\documentclass[3p,dvipsnames,preprint]{elsarticle}
\usepackage{algorithm}
\usepackage{bbm}
\usepackage{algpseudocode}
\usepackage{multirow}
\usepackage{makecell}
\usepackage{footnote}
\usepackage{tikz}
\usepackage{eso-pic}
\usepackage[caption=false,font=footnotesize]{subfig}
\DeclareGraphicsExtensions{.pdf,.png,.jpg,.eps}

\usepackage[final]{changes}

\usepackage{ulem}

\definechangesauthor[name={Andrea Passarella}, color={red}]{AP}
\definechangesauthor[name={Lorenzo Valerio}, color={blue}]{LV}
\definechangesauthor[name={Matteo Mordacchini}, color={OliveGreen}]{MM}
\usepackage{soul}
\sethlcolor{white}

\begin{document} \begin{frontmatter}
\markboth{M. Mordacchini et al.}{A semantic and cognitive-based solution
for  knowledge and content spreading in OppNets}
%
%\title{A cognitive-based semantic-aware solution for knowledge and content
%  spreading in opportunistic networks}
\title{Design and evaluation of a
cognitive approach for disseminating semantic knowledge and content in
opportunistic networks}

\author[cnr]{Matteo Mordacchini} \ead{m.mordacchini@iit.cnr.it}
\author[cnr]{Lorenzo Valerio\corref{cor1}} \ead{l.valerio@iit.cnr.it}
\author[cnr]{Marco Conti} \ead{m.conti@iit.cnr.it}
\author[cnr]{A. Passarella} \ead{a.passarella@iit.cnr.it}
\cortext[cor1]{Corresponding author}
\address[cnr]{Institute for Informatics and Telematics, National Research
Council. Via G. Moruzzi 1, 56214 Pisa, Italy}
\begin{abstract}
  In cyber-physical convergence scenarios information flows
  seamlessly between the physical and the cyber worlds. Here, users' mobile
  devices represent a natural bridge through which users process acquired
  information and perform actions. The sheer amount of data available in this
  context calls for novel, autonomous and lightweight data-filtering solutions,
  where only relevant information is finally presented to users. Moreover, in
  many real-world scenarios data is not categorised in predefined topics, but it
  is generally accompanied by semantic descriptions possibly describing
  users'   interests. In these complex conditions, user devices should autonomously
  become aware not only of the existence of data in the network, but also of
  their  semantic descriptions and correlations between them. To tackle these
  issues, we present a set of algorithms for knowledge and data dissemination in
  Opportunistic Networks, based on simple and very effective models (called
  cognitive heuristics) coming from cognitive sciences. We show how to exploit
  them to disseminate both semantic data and the corresponding data items. We
  provide a thorough performance analysis, under various different conditions
  comparing our results against non-cognitive solutions. Simulation results
  demonstrate the superior performance of our solution towards a more effective
  semantic knowledge acquisition and representation, and a more tailored content
  acquisition.
\end{abstract}
%\begin{bottomstuff} This work is funded by the EC under the %FET AWARENESS
%RECOGNITION (FP7-257756), FIRE EINS (FP7-288021) project.\\ Author's addresses:
%M. Mordacchini, matteo.mordacchini@iit.cnr.it \end{bottomstuff}
%
%\maketitle
%
\end{frontmatter}
	
\section{Introduction}
\label{sec:Intro}

The physical world is becoming more and more saturated by the  presence of a
vast number of mobile devices. These devices are able to sense data from the
physical environment and autonomously elaborate and exchange information among
themselves. This behaviour is leading to a constant and increasing flow of
information from the physical world to the cyber one and vice-versa.  In this
context, data coming from the physical world impacts on the decisions taken by
devices acting in the cyber world, whereas the information spread in the cyber
world can, in turn, influence the actions taken in the physical world. This
complex information scenario is known as the {\em Cyber--Physical World} (CPW)
convergence
scenario~\cite{CPW,Sahoo2014,Park:2012aa,Borgia:2014aa,Conti20112115,Al-Hammouri:2012aa}.

The users' devices acting in this scenario play a key role, since they represent
the ``door'' to the cyber world through which their human owners can access the
massive amount of information available in the cyber world. 
%MATTEO TODO COMMENTO ANDREA SU BANDWIDTH CRUNCH DA REPORT CISCO since they are
%one of the most important ways for their human owners to access to  the massive
%amount of information available in the cyber world.  In this way, devices in
%the CPW convergence scenario are regarded as the {\em avatars} of their users
%in the cyber world. Their role in the discovery and retrieval of relevant
%information present in the cyber world environment is so crucial that
%particular care should be devoted in designing proper algorithms for those
%sensitive tasks. 
Indeed, in the CPW convergence scenario users'
personal devices can
be regarded as the \emph{proxies}
of their users in the cyber world. They are most of the time together
with their users in the physical world, and can thus, for example, gather
context information about their behaviour and the physical places they visit. On
the other hand, they are probably the most typical way through which users access
information in the cyber world. Therefore, they can be usefully instructed to
autonomously act in the cyber world (e.g., by proactively filtering or fetching
information) on behalf of their users, by exploiting context information about
their behaviour in the physical world.
To this end, because the cyber world is typically full of data of all
kinds that could possibly be accessed, users' devices should automatically
understand which data is important for their users at a given point in time,
avoiding to overflow them with useless information.
%Moreover, networking issues aside, in such extremely data rich mobile
%environments, users' devices must be able to automatically understand which
%data is important to their users at a given point in time. Therefore, they can
%be regarded as avatars...

%In previous works (e.g.~\cite{TAAS12}), we have emphasized the key role played
%by the users' devices acting in this scenario. In fact, they are one of the
%most profitable mean allowing their human counterparts to access the massive
%amount of information potentially available in the cyber world. In this way,
%devices in the CPW convergence scenario will be regarded as the {\em avatars}
%of their users in the cyber world. Their role in the discovery and retrieval of
%relevant information present in the cyber world environment is so crucial that
%particular care should be devoted in designing proper algorithms for those
%sensitive tasks. These algorithms should let mobile devices rapidly and
%efficiently recognise the relevance of the data they discover during encounters
%with other nodes. In order to properly fulfil this task, such solutions should
%let the devices to analyse the nature and features of the data they are exposed
%to. 

 In this context, the opportunistic networking paradigm plays a relevant role by
 supporting direct communication between mobile devices. In an opportunistic
 network, direct, physical contacts between nodes are opportunistically
 exploited to recognise and disseminate relevant information toward potentially
 interested nodes, without the need of centralised infrastructures or
 precomputed paths from source to
 destination~\cite{Pelusi:2006th,Ferretti:2013aa,Mota:2014aa,Boldrini:2014ab}. Beyond the problem of data dissemination, it is worth to mention that the other main research issues in opportunistic networks focus on the development of
analytical models of data delivery
performance~\cite{Boldrini:2014aa,Sermpezis:2014aa,Ginzboorg:2014aa}, routing
approaches that consider nodes' aggregation and privacy
~\cite{Chen:2014aa,Aviv:2014aa}, forwarding schemes for hybrid networks
(opportunistic networks combined with the
infrastructure~\cite{Mayer:2014aa}), real world implementations
~\cite{Grasic:2014aa}, applications to Vehicular Networks~\cite{Benamar:2014aa,Kokolaki:2014aa}. Recently,
 opportunistic networks have been proposed as one of the possible key components of
 future mobile networks (e.g., in the 5G domain), as they are able to complement
 wireless infrastructures such as cellular and WiFi networks, by enabling direct
 dissemination of data among users nearby, thus contributing to offload data
 from the cellular network~\cite{Rebecchi:2014ys,Conti:2014aa}.

%In order to optimise the diffusion of data in this scenario, we Specifically,
%consider that in many real-world scenarios data items (contents) are equipped
%with proper semantic descriptions (i.e. associated concepts and/or tags). This
%could be the case, for instance, of tagged photos on Flickr and Instagram, or
%messages annotated with ``hashtags'' in Twitter and Facebook. In those
%contexts, users' interests toward contents are driven by the data items
%descriptions.
 Generally, in analysing data dissemination in opportunistic networks  it is
 assumed that the users' interest is rather
 static and quite simple to
 describe~\cite{Boldrini10}. In typical data
 dissemination approaches, users are supposed to be
 interested in predefined content categories (e.g., sports, movies, etc.) and
therefore their devices collect all the
contents related to those categories.
An aspect that has not been often taken into consideration in the literature on
opportunistic networks is that contents are also equipped
with rich semantic
descriptions (i.e. associated concepts and/or tags). This could be the case, for
instance, of tagged photos on Flickr and Instagram, or messages annotated with
``hashtags'' in Twitter and Facebook. Often, users' interests toward contents
are driven by the content descriptions , which triggers interest in contents themselves: data items\footnote{In the paper we use the terms content and data item interchangeably.}  are accessed by users
because their semantic description contains information that is relevant to the
user at that moment. Therefore, our key idea in this paper is to optimise  in an intertwined way the dissemination of both \emph{semantic} data associated to
  contents, and of \emph{contents} themselves. Specifically, in the proposed
  approach semantic data is disseminated among users' devices, based on the
  current interests of the users. By receiving semantic data, users' devices
  also know what contents (associated to that semantic data) are available in the
  network, and fetch them, if appropriate. Therefore, in our approach users'
  (dynamic) interests drive the dissemination of semantic data, which drives the 
  dissemination of content. As we explain in Section~\ref{sec:CP}, this
  mechanism is quite similar to the way users access information in the physical
world.

% In order to effectively filter the data found in the environment and select
% the most useful content for both their users and the overall content
% dissemination in the network, devices should be able to get aware of the main
% features that describe the contents spread in the surrounding environment.
% This should be done considering how these descriptions can be linked together
% and how they can be related to the users' interest in order to select and
% disseminate only the most relevant content.

%~\cite{Mordac11}

Specifically, in this paper we  apply concepts
coming  from the cognitive science field, and design a system whereby users'
devices autonomously become aware of
the structure of the semantic information describing the available content in
the environment, and disseminate contents based on this knowledge.  In particular, we show how devices can
exchange information in a way that resembles how conversations between humans
enable spreading of ideas (i.e. semantic information), which generates interests
for specific types of content, and ultimately determine content that people
access.  To this end, we consider that, acting on behalf of their users inside
the cyber world, mobile devices are exposed to problems similar to the ones
faced by the human brain when dealing with the information and content selection
tasks. Cognitive scientists produced, during years, many functional models and
descriptions of these mental schemes. These functional models, called {\em
cognitive heuristics}, differently from other biological models as artificial
neural networks, do not aim at reproducing the physiology of the brain's
processes, but model their functionality. By taking advantage of these
descriptions, previous works~\cite{TAAS12,Valerio:2014qf,Bruno12}  have shown
how rules and procedures used by the human brain, when assessing the relevance
of 
information (in face of time and resource restrictions), can be exploited to
design adaptive, low resource-demanding, yet very effective, algorithms for
data dissemination in opportunistic networks but none - to the best
of our knowledge - has explored how to apply these models to optimise the joint
dissemination of semantic information and associated contents.

 In order to take advantage of these models, we have to face the problem of how
 to represent semantic information in devices' memory, how semantic
 information is retrieved and exchanged upon contacts between nodes in physical proximity, and how content is finally selected for dissemination,
 based on the semantic data exchange that
 has been carried on.
For each node, the internal memory representation of semantic concepts is
inspired by the associative network models
(AN)~\cite{Anderson:1979sf,Collins:1975vn} of human memory coming the from
cognitive psychology field. In AN models, semantic concepts are represented by
nodes that are interconnected by paths that vary in strength, reflecting the
degree of association between each pair of concepts. In our proposal, each
mobile node builds a local semantic representation of its own contents through a
semantic directed weighted graph, where vertices represent the semantic concepts
associated to data items, and the edges represent the semantic relationships
between concepts, both learnt from the environment and derived from the node's
own contents.  Moreover, since memory is a limited resource at each node, cognitive
models of how the least relevant information can be dropped from
memory~\cite{Wixted:1991zl,Ebbinghaus13} are exploited. When nodes come
  in physical proximity (are in contact), exchange of
semantic information happens as follows.
Communication takes place only between nodes having some common interest,
i.e. only if there is an initial set of semantic concepts shared by both
nodes. Concepts to be exchanged are
selected by navigating each semantic network, according to an edge ranking
algorithm derived from the {\em fluency } cognitive heuristic
(FH)~\cite{Schooler05,Marewski:2011ul}.  Finally, we show how mobile nodes, after having enriched their
semantic graphs with new concepts taken from other encountered nodes, select
which contents, locally available at one of the nodes, to exchange
between them, giving precedence to contents whose semantic information maximally
overlaps with semantic concepts just exchanged between nodes. In cognitive
terms this refers to the {\em tallying heuristic} (TH)~\cite{Dawes79}, another
cognitive decision strategy used by human brain.  All these cognitive processes
are described in more details in Section~\ref{sec:CP}.

% MATTEO TODO : mettere alcuni valori numerici particolarmente significativi. 

The rest of the paper is organised as follows. In Section \ref{sec:RW} we
present state-of-the-art data dissemination approaches for opportunistic
networks. In Section \ref{sec:CP} we show at a high level how each cognitive
model relates to our solution. In section \ref{sec:approccio} we present the
entire approach in full detail. In Section \ref{sec:expRes} a thorough
performance analysis of our cognitive based system is provided and, finally,
Section \ref{sec:conclusion} concludes the paper. 
% SPOSTATO QUI DA SEZIONE 3
%Note that, with respect to the popular topic based approaches described above,
%the use of semantic networks to represent concepts that are then linked to the
%data tags can be seen as a generalisation of the use of topics to categorise
%data, much closer to the way humans categorise information in their memory.
%Moreover, due to the characteristics of these cognitive processes (\emph{fast
%and frugal}~\cite{Giger08}) it is possible to design solutions characterised by
%a low computational complexity. 

%:- Related work
\section{Related data dissemination approaches}
\label{sec:RW}

The problem of data dissemination in opportunistic networks have been
addressed by many works during years. The first attempt was in the context of
the PodNet Project~\cite{Lenders08}. The authors proposed a solution where nodes
cooperatively exchange data items in order to retrieve all those contents they
are interested in. Precisely, contents are organised in predetermined channels
of interest to which nodes are subscribed. In order to favour the
data
dissemination, upon encounter, nodes load in a \emph{public cache} items they
are not directly interested in. Items to be maintained in the public cache are
chosen depending on different history-based strategies that consider the past
received requests, interpreted by nodes as a popularity index of the channel of
interest. These strategies could be effective when users mobility is homogeneous
and contents can easily traverse the network.  However, this approach suffers in
scenarios where nodes tend to group in communities and their movements are
heterogeneous.
%~\cite{Boldrini:2008dk,Boldrini:2008la

Advances in data dissemination solutions leave the content-centric approach
adopted in PodNet in favour of more user-centric solutions. Precisely, these
solutions define more elaborate heuristics that exploit social information about
users -- e.g. users' interests and social relationships  --  in order to improve
the performance of data dissemination.  ContentPlace~\cite{Chiara10} is a
social aware dissemination system that proposes a general framework for
designing data dissemination policies. In ContentPlace,
during a contact nodes fill their local caches (with contents they either
already have or they fetch from the other node) to maximise both the local utility (i.e. satisfy the interests of
the local  user) and the global utility (i.e. satisfy the
interests of other nodes in
the local user's communities). %FORSE AGGIUNGERE QUALCOSA D'ALTRO PER CONCLUDERE
In \cite{Pantazopoulos:2010rw} authors propose a social-aware
solution to find the optimal placement of a given piece of content in an
opportunistic network. The idea is to iteratively migrate contents to nodes that
are increasingly ``central'' to the overall network, i.e. nodes such that the
average cost of accessing the content from any other interested node is
increasingly lower.  Social-aware data dissemination
generally results in a quicker and more fair content dissemination with respect
to social-oblivious policies. The overhead introduced for collecting
and managing the information needed by these algorithms (e.g., contact patterns,
social structures, etc.) typically pays off as content can be replicated less
aggressively (and more precisely), and the total network traffic typically turns
out to be lower than in social-oblivious schemes.

Other works apply the publish/subscribe framework -- a content-centric overlay
initially conceived to run on top of static networks~\cite{Eugster:2003rz} -- to
opportunistic networks, by adapting the conventional definitions of
publisher, subscriber and broker.
Accordingly, in \cite{Yoneki07}  the authors presented a pub/sub
social-based and topic-based data dissemination solution. Brokers are the most
central (i.e. popular) nodes in the community and communities are identified
through an on-line community detection algorithm. New data items to be diffused are
sent to the broker of the community. If some node is subscribed to that topic
within the community, the item is broadcast. Moreover, the item is also sent to
other brokers if their community members are subscribed to that topic.

Other approaches  take in account information about
patterns of encounters. In SocialCast \cite{Mascolo07} nodes disseminate
the information about their channels of interest. Every node uses this
information combined with its own pattern of encounters to compute a utility
value for each channel of interest. This value is then used to decide if the
next encountered node would be a good carrier for a given data item.   In
PrefCast \cite{Lin:2012yq} authors assume that the utility of disseminating the
same content varies with time. Considering limited contacts
durations, nodes compute a forwarding schedule to prioritise dissemination of
contents that are predicted to be more requested in the near future.
	
Other approaches for data dissemination consider solutions based on
global utility functions to be solved as a global optimisation
problem, where nodes'
individual caches are viewed as a big, cumulative caching space.  Reich and
Chaintreau~\cite{Reich:2009oz}, for example, focus on the problem of finding a
global optimal allocation for a set of content items assuming that users are
impatient, i.e., users' interest for items monotonically decreases with
 time. Although such global optimisation approach can find
the optimal policy and lead to an optimal content allocation,
it requires global knowledge of the network and a-priori information
on how users behave, that in practice, might be unlikely to be available in an
opportunistic scenario.

%\cite{Bruno12
The use of cognitive heuristics to drive the dissemination process has
  been first proposed in~\cite{TAAS12}. The idea is to drive the dissemination
process using procedures that mimic, in functional terms, the human decision
making  process. To this end, two of the several cognitive models present in the
psychological literature are considered~\cite{Giger08}:  the \emph{Recognition
Heuristic} and the \emph{Take the Best Heuristic}. Precisely, nodes use the
Recognition heuristic to become aware of which are the most popular channels of
interest in the system and what are the corresponding contents that should be
further replicated in order to satisfy all the interested users. This work
proves  the suitability and effectiveness  of these heuristics in problems, like
data dissemination in opportunistic networks, where every node
has only a partial knowledge
about its environment. However, due to the great number of status information
the nodes have to maintain, this approach may suffer from scalability problems. 
%Valerio13
In \cite{Valerio:2014qf} an improved solution  addressed those scalability
problems. Nodes exploit, beyond cognitive heuristics, i) a local aggregate
measure about items diffusion and ii) a stochastic mechanism to choose which
data items should be replicated. The resulting solution shows to be more
efficient and scalable than the previous one and, thanks to the stochastic
decision making mechanism, independent of specific scenario configurations. 

Another different but connected solution presented in the literature takes into
account the semantic side of data shared in an opportunistic
network~\cite{Saso13}.  Precisely, starting from semantic data annotations given
by the users, each node builds a  semantic network representing that
information. Upon contacts, nodes exchange useful information present in their
semantic networks which is thus disseminated in the entire network. That work
represents the first attempt to exchange semantic information exploiting
cognitive mechanisms in opportunistic networks.

In this paper (which extends our previous work~\cite{Aoc13}) the
ideas of~\cite{Saso13}
are exploited to design a more complete cognitive-based
solution for both knowledge (semantic information) and content dissemination.
The main extensions with respect to~\cite{Aoc13} include a more detailed description of all
the cognitive models and concepts exploited in this proposal, a more thorough
presentation of the algorithms developed from the cognitive science models, and
a more extensive set of simulation results. These results analyse the performance of
the system from both the semantic knowledge and the content dissemination
points of view, using different metrics and under different settings of the main
parameters of the system. Moreover, we compared the proposed solution with
another algorithm for semantic and content dissemination that is not based on
cognitive models. We find that the proposed solution is much more efficient in
retrieving and keeping in memory the most relevant semantic information and it
acquires the relevant content associated with this information more rapidly than
the benchmark solution.

We point out that part of the semantic-driven exchange of information
  process described in this paper is also used in~\cite{TAAS15}. Specifically,
  in~\cite{TAAS15} cognitive-based methods are used to exchange semantic
  information, in order to let mobile nodes become aware of the features that
  describe the physical locations in the environment around them. Semantic data
  is generated by the physical locations and it is spread by both physical
  locations and mobile devices. With respect to that work, in this paper we
  exploit the opportunistic semantic knowledge dissemination mechanism for
  different purposes and under different conditions. Specifically, in this work
  we use a semantic exchange of information in a pure opportunistic network,
  i.e. contacts between mobile devices are the {\em only way} to exchange and
  disseminate information. Content is generated and spread only by the devices. 
  These facts determine a different behaviour of the two systems and
  require changes between the two solutions, specifically in the definition of
  the key algorithms of the proposed approach
  (see Sec.~\ref{sec:fluency}). Moreover, in this
paper we do not propose a simple semantic data dissemination scheme. Rather,
semantic information is used to drive the spreading of the data items from which
the semantic descriptions are taken. In order to achieve this goal, other
mechanisms are needed in addition to the semantic dissemination scheme. To this
end, we propose to exploit another lightweight, context-aware, cognitive-based
process for the dissemination of data items, which is driven by the
dissemination of the associated semantic data. 

%We point out that part of the semantic-driven exchange of information process
%described in this paper is also used in~\cite{Mordacchini:2015aa}.
%Specifically, in~\cite{Mordacchini:2015aa} cognitive-based methods are used to
%exchange semantic information, in order to let mobile nodes become aware of the
%features that describe the physical locations in the environment around them.
%Semantic data is generated by the physical location and it is spread by both
%physical locations and mobile devices. With respect to that work, in this paper
%we exploit the opportunistic semantic knowledge dissemination mechanism for
%different purposes and under different conditions. Specifically, in this work
%we use a semantic exchange of information in a pure opportunistic network, i.e.
%contacts between mobile devices are the {\em only way} to exchange and
%disseminate information. Content is generated and spread only by the devices
%themselves. These facts determine a different behaviour of the two systems and
%requires changes between the two solutions, specifically in the definition of
%Algorithm~\ref{alg:visit} (see Sec.~\ref{sec:fluency}). Moreover, in this paper
%we do not propose a simple semantic data dissemination scheme. Rather, semantic
%information is used to drive the spreading of the data items from which the
%semantic descriptions are taken. In order to achieve this goal, other
%mechanisms are needed in addition to the semantic dissemination scheme. To this
%end, we propose to exploit another lightweight, context-aware, cognitive-based
%process.

\section{High-level overview}
\label{sec:CP}

In this section we provide a high
level overview of our approach. Namely we provide a direct mapping between the
aspects of memory representation, concept retrieval and content selection, and
the cognitive models we exploit to define the building blocks of our approach. 
	
%:-- Associative Semantic Network
\subsection{Memory Representation}
\label{ssec:memory}

The first aspect of our
solution concerns the representation and organisation of the semantic
information about content in nodes' memory. The organisation of the semantic
information is a very common task that the human memory solves very well. Thus,
 we used  one of the
memory models present in the cognitive science field. 
Recently, two categories of models are significantly well established in
the literature, namely, the Associative
Network Models and the Connectionist Models~\cite{Anderson:1979sf,Collins:1975vn}. 

In \emph{Associative Network Models}, concepts are represented by nodes that are
interconnected by pathways that vary in strength, reflecting the degree of
association between each pair of concepts. Two related concepts are connected by
links, whose weight represents the activation level, i.e. how likely it is that
one of them is ``accessed" after the other has been accessed. The details of how
these weights are computed are provided in Section \ref{sec:memrepr}.
		
\emph{Connectionist Models}  represent an alternative to Associative Network
Models.  These models treat the problem of mental representation as weighted
combination of fixed set of features~\cite{Rumelhart86}. Though the
Connectionist models share some terminology and ideas with the Associative
Network Models, they follow a very different metaphor. Precisely, they consider
the memory as a large network of feature units (nodes) that share activation
through weighted connections (pathways). Therefore, in these kind of networks,
concepts representation are distributed across the entire network
and this generates a process
that is very similar to what happens in biological neural networks. Indeed, the
activation of units is affected  by the weights of the incoming links that  can
be either positive or negative. Such models are more sophisticated than the
associative network models described before, but they are also more complex,
i.e. in order to converge to a good representation of concepts they must be
trained and the training process is typically very slow. 

\begin{figure}[ht] \centering
  \includegraphics[width=.65\textwidth]{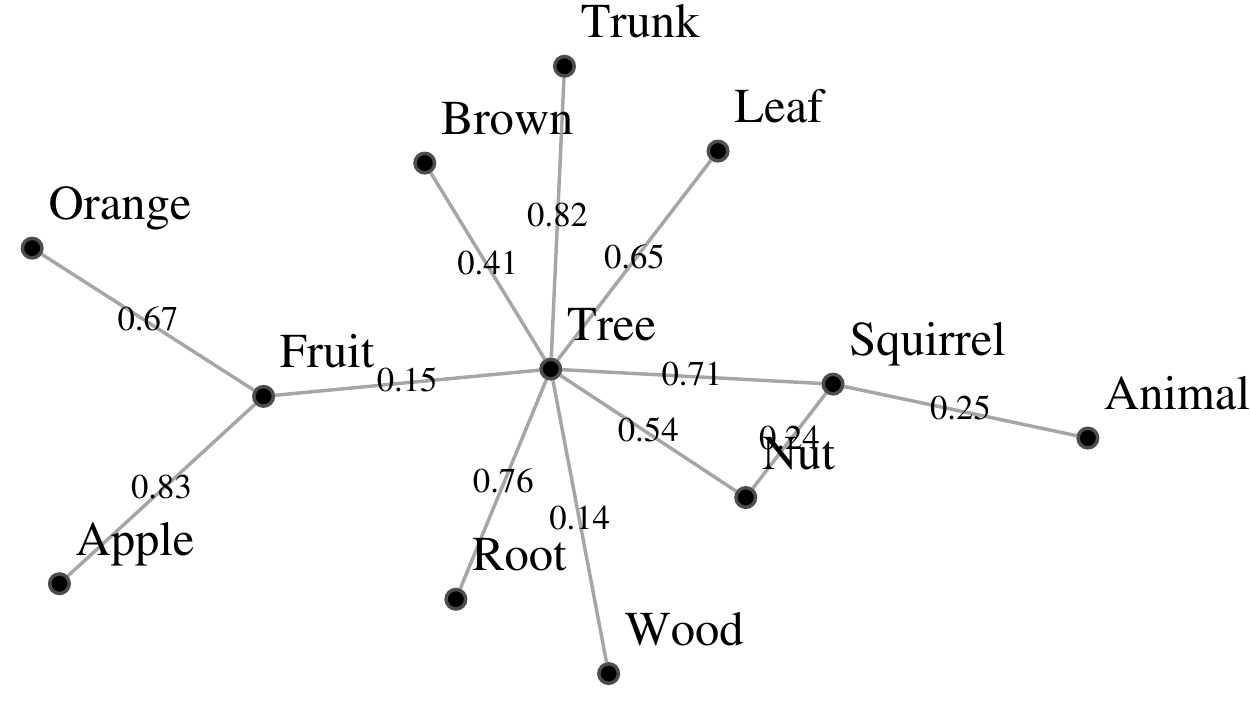} \caption{Example of
  associative semantic network} \label{fig:asn} \end{figure}

In our work we use Associative Network Models and precisely Associative Semantic
Networks due to their simplicity and usability. In our solution nodes of the ASN
represent semantic concepts describing the content in the system. Nodes
relations are represented by weighted edges. In Fig. \ref{fig:asn} an example of
Associative Semantic Network is reported.  

\subsubsection{Memory retention}
\label{sec:memret} 

\begin{figure}
  \centering
  \includegraphics{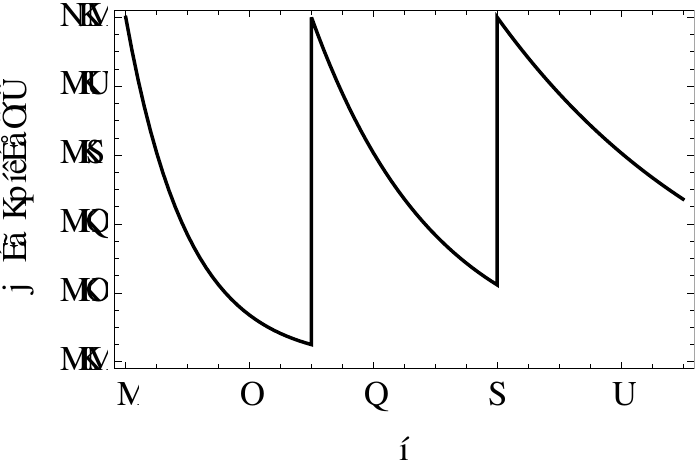}
  \caption{Human memory strengthening process according to the H. Ebbinghaus' model.
    \label{fig:forgetHeb}}
\end{figure}

In order to be more compliant
with human memory models the Associative Semantic Networks we used are
dynamic. This means that the structure of the ASN can change over time depending
on its use. In fact, like in  human memory, an information that is not accessed
for some time fades in our memory. Conversely, the more we access the same
information the stronger it remains in memory. The process of forgetting useless
information from memory, rather than a limit, aids the brain in improving the
effectiveness of some human cognitive processes \cite{Schooler05} by letting them
keep and consider only relevant information.  In particular, we model
the
forgetting process in our system by a ``forgetting'' function that is similar to
an experimental forgetting curve early obtained by H. Ebbinghaus in 1885
\cite{Ebbinghaus13} and described by numerous subsequent studies
\cite{Wixted:1991zl}. This curve models the human forgetting process as an
exponential decline of individual human memory retention over time, where the
rate of forgetting is determined by the repetition rate of information (i.e. by
the process through which that information is accessed in the brain). Fig.
\ref{fig:forgetHeb} shows an example of this type of curve. The slope of the
forgetting curve slows down more gradually whenever a stored information is
accessed or observed more frequently. Thus, we can assume that unused
information is easily forgotten as time passes, while frequently accessed
information becomes harder to forget.  In the case of Associative Semantic
Networks this mechanism affects the strength of edges linking semantic concepts,
determining, as a consequence, the evolution of the ASN's structure. Further
details on how we implemented this model in our system will be given in
 Section~\ref{sec:approccio}.

\subsection{Concept retrieval}
\label{ssec:fluency}

The second aspect is about
the way the semantic information (knowledge) circulates in the network. Here
also, we exploit mechanisms coming from the cognitive science field and
resembling the process  of information exchange between people during a
conversation. We model information exchange when nodes encounter as a
conversation between two individuals who meet. We assume that the conversation
begins from concepts that people
have in common -- i.e. concepts present in all participants semantic networks --
and that the exchange of semantic concepts consists in the
exploration/navigation of the reciprocal semantic networks. There are two
principal ways to explore an Associative Semantic Network:
sequentially or in parallel, which broadly corresponds to  depth-first and
breadth-first navigation in standard graph theory,
respectively~\cite{Gawronski}.  	 Parallel processing is particularly
appropriate to model the spontaneous activation of knowledge that occurs in
absence of a particular goal. Sequential search, instead,  well characterises a
goal-directed search of memory in pursuit of a specified objective.  A
sequential search (SS) on AN starts with an activated concept (key-concept) and
proceeds node by node along the pathway that connects them. When a node has many
outgoing paths, the one with the {\em strongest activation} (weight) is
selected. If a ``dead end" is reached, the search is re-initiated. Here, we
concentrate on the sequential search model (SS) due to its better pertinence to our
purposes. In fact, in our model each device, starting from a commonly shared
knowledge (the ``base of discussion", in the  conversation metaphor), uses an information searching mechanism to retrieve
from its memory the most semantically correlated  data  to pass to another
interacting party. This goal-oriented information selection process can be
better described with the SS model rather than with a parallel search.

 % MATTEO TODO spiegare perch��� ci focalizziamo sulla SS (ad esempio dicendo
 % che vogliamo simulare una flusso di concetti strettamente connessi tra loro
 % al posto di saltare di palo in frasca. 

 The strongest activation rule is the implementation in this context of the
 first {\em cognitive heuristic} that we use in the paper, i.e. the
 \emph{Fluency Heuristic} (FH).  
%	Concept retrieval is accomplished by means of the \emph{Fluency
%	Heuristic} (FH). 
FH is an inference strategy that can be applied when someone has to choose among
two or more alternatives. Among the alternatives that are \emph{recognized}, the
one perceived as recognized faster is considered to be more important
with respect to a
selected criterion. Here, being recognized means that a given information has
been found in the environment a sufficient number of times to let the brain
being familiar with it .
%Thus, if both alternatives can be retrieved from memory, the quickest retrieved
%is considered  more appropriate. 
In the context of the navigation of a semantic network from a given concept, the
fluency heuristic dictates that the next concept is the one linked through the
edge with the highest weight, because this is what allows the brain to retrieve
it faster from memory.
% !! TOLTO L'ESEMPIO, MA VOLENDO SI PUO' RIMETTERE
%To give an example, let us recall the  conversation metaphor presented in
%Sec.\ref{sec:Intro} and consider  Figure \ref{fig:esempioFH}. Suppose that the
%current concept in a conversation is ``Tree'' and a person needs to select the
%next proper concept to continue the conversation. As we can see, the ``Tree"
%node has many other concept attached, some of them are \emph{recognized} (green
%nodes) and some others are not recognized (yellow nodes). Among the recognized
%ones, FH selects the concept with the strongest activation  ``Fruit" (edge's
%weight value is $0.8$).      \begin{figure}[H] \centering
%\includegraphics[width=.75\textwidth]{img/esempioFluency} \caption{Example of
%fluency heuristic in action } \label{fig:esempioFH} \end{figure}
The details of how the fluency heuristic is applied to recognise
concepts, and identify those
that are exchanged between nodes are presented in Section \ref{sec:fluency}.

%:-- Tallying Heuristic
\subsection{Content selection}
\label{ssec:tallying}
The third aspect of our system is the actual data exchange. Here we exploit the
mechanism of information exchange during a conversation. Based on the concepts
exchanged during the phase sketched in section \ref{ssec:fluency}, content items
are exchanged giving preference to those  that are more ``central''
with respect to the topics just discussed (i.e. concepts just
exchanged). This mechanism is modelled, in cognitive terms, by the
\emph{Tallying Heuristic} (TH). Precisely,  in order to discriminate among
alternatives, TH selects those alternatives that according to a criterion, have
more favourable cues. In our case, cues are tags associated to data items, and a
tag is favourable if it matches one of the concepts accessed in their semantic
networks by two nodes during an encounter (i.e., if the nodes have ``spoken"
about a concept that matches the tag).  Precisely, for each alternative, TH
simply counts the number of favourable cues without giving any special weight to
any of them. The alternative with the highest number of positive cues is then
selected~\cite{Dawes79}.  Tallying has proved to perform the same or even better
than multiple regression models. As an example, let us consider a conversation
between two users, \emph{A} and \emph{B}, on how beautiful  mountains
are especially during winter. At the end of their conversation, user \emph{A}
remembers that some pictures stored in his smartphone are about his last holiday
spent on the Alps. Thus, user {\em A} selects those pictures, and shares them
with user {\em B}. In this example, according to the Tallying Heuristic, the
pictures are selected by user {\em A} because they are the ones
having more favourable cues with respect to all the other
pictures stored in {\em A}'s phone. 
% MATTEO TODO: fai un check all'esempio che ho messo sulla tallying (marco
% chiedeva un esempio per chiarire il concetto)

%\begin{figure}[hb] \centering
%\includegraphics[width=.7\textwidth]{img/esempioTallying1.pdf} \caption{Expamle
%of the tallying heuristic in action} \label{fig:esempioTallying} \end{figure}
%In this work, we use these three cognitive building blocks to disseminate both
%knowledge and contents in opportunistic networks. 
Details on how this is implemented in our scheme are provided in Section
\ref{sec:tallying}	

%:- Cognitive heuristics for data and content distribution in OppNets

\section{Cognitive heuristics for data and content distribution in
  Opportunistic Networks}
\label{sec:approccio}

In this section, we describe in detail how cognitive models are used to define
the building blocks of our data dissemination system. In Sections
\ref{sec:memrepr} and \ref{sec:forget} we describe how semantic information
(concepts) are stored in memory and dropped. Section
\ref{sec:fluency} describes how we use the fluency heuristic to decide which
concepts to exchange between nodes upon encounter. Finally Section
\ref{sec:tallying} describes how we select data items to exchange based on the
tallying heuristic. 

\subsection{Memory Representation of  Semantic Concepts}
\label{sec:memrepr}

%
%In order to obtain a useful description of the semantic knowledge and the
%connections between different semantic concepts that are spread in the
%environment, each node need to use a proper model that let it organize the
%acquired semantic knowledge in its memory. 

As anticipated in Section~\ref{ssec:memory}, one way used by cognitive
scientists to describe how humans store their semantic information in memory is
given by the {\em Associative Network} models. In these models, semantic
concepts are viewed as nodes and a couple of nodes is connected whenever the
brain is able to establish a relationship between the involved semantic
concepts. These relationships could have different strengths that vary over
time, reflecting the intensity with which the brain perceives those association.

In order to let mobile nodes exploit this cognitive-based memory organisation,
we define each user's semantic network as a {\em dynamic undirected weighted
graph}  $G = \{V,E,f(e,t)\}:t \in T$ , where: \begin{itemize} \item  $t$ is the
      time at which the graph $G$ is evaluated \item $V$ is the set of vertices,
      i.e. the semantic concepts known by the node  \item $E$ is the set of
	edges, i.e. the connections between semantic concepts the node is able
	to recall at time $t$ \item $f(e,t)$ is the function that defines the
	  weights of each edge $e \in E$ at time $t$ \end{itemize} 

In a scenario where users actively participate in the creation of  content, we
assume that each data item owned by a user is associated to a set of tags --
defined by the user herself -- that semantically describe it. This process is
similar to what happens in real social networks like Flickr, Twitter, Instagram,
etc.  The evolution of the graph will be clear after explaining the algorithms
in the following sections. For now, it is important to explain how the graph is
initialised at each node at time $t_0$, which is assumed as the beginning of the
evolution of the system\footnote{Without loss of generality, we assume that all
nodes start at time $t_0$. Algorithms do not change when nodes enter the system
at different points in time.}.   For any given user, at time $t_0$, its
semantic network defined by $G$ includes all the
data tags associated with the data items locally available at the node, as its
vertices. The set $E$ of the edges of $G$ is created using the following
algorithm. Taking one data item at a time, the tags of the data item are linked
together in order to form a completely connected component. After that, the
vertices that belong to different components and have the same label  (i.e. they
were created from tags having the same name) are merged together, forming a
single vertex in the final graph $G$. This single vertex inherits all the
ingoing and outgoing edges pointing to the original vertices in their respective
connected components.

\begin{figure}[ht]
  \centering \includegraphics[width=0.6\columnwidth]{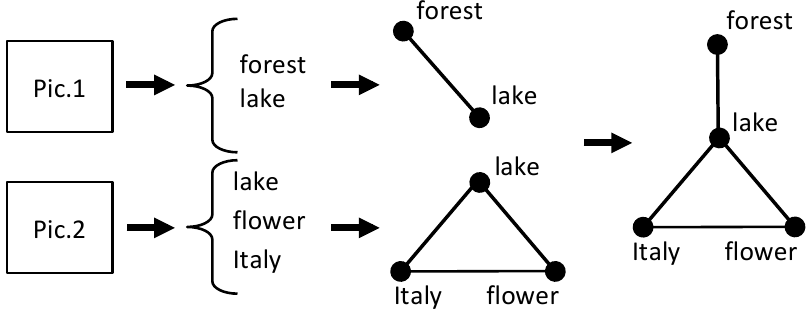}
  \caption{Creation of a user Semantic Network at time $t_0$.}
  \label{fig:fusion} \end{figure}

In order to better clarify this process, Fig.~\ref{fig:fusion} presents an
example of the mechanism described above. In this example, we assume that there
is a user with two different pictures. Each of the pictures has a set of
associated tags. Firstly, two completely connected components are created, one
for each of the two pictures. Then, the two components are merged together using
the common vertex ``lake" as the pivot of this process.  

%In Sec.~\ref{}, we stated that the semantic concepts of a node and the
%relations between them are described using an associative network model. More
%formally,  each user's semantic network is defined as a {\em dynamic weighted
%graph} $G = \{V,E,f(e,t)\}:t \in T$ , where $t$ is the time, $V$ is the set of
%vertices (i.e. semantic concepts) in the graph and $E$ is the set of edges
%(i.e. the connections between semantic concepts).

\subsection{Memory Retention}
\label{sec:forget}
%%%%%%%%%%%%%%%%%%%%%%%%%%%%%%%%%%%%%%%%%
%
%The description of the users' semantic memory organisation allows us to define
%how the semantic knowledge is stored by each node and how the initial
%configuration  of this memory is derived from the data created by each device.
%However, in Sec.~\ref{sec:memrepr}, we stated that the graph representing a
%node's semantic network is {\em dynamic}.
 One of the sources of changes in the nodes' semantic network, that make it
 dynamic, is the loss of information, that we model with a {\em forgetting
 process}, as follows. 
% One source of changes is the potential acquisition of new knowledge received
% during encounters with other nodes. Another relevant mechanism that leads to
% changes over time in the semantic network of a node is the deletion of
% vertices and edges as a result of a {\em forgetting} process.
The rationale behind the forgetting process is twofold. On the one hand, it
takes into consideration memory limitations at nodes. On the other hand, it
mimics the typical behaviour of the human brain, that forgets information that
is not accessed for some time. Even disregarding physical memory limitations of
devices, such a forgetting process is an automatic way of assessing the
relevance of information for the user, based on how often the user accesses it. 
Exploiting the cognitive definition~\cite{Ebbinghaus13,Wixted:1991zl} of the
human forgetting function (see Section~\ref{sec:memret}),  we can define the
forgetting function for each edge $e_{ij} \in E$ of a user's semantic network at
a time $t$ as: 

\begin{equation} f(e_{ij}, t) = e^{-\beta_{ij}(t-t^*)} \end{equation}

\noindent  where $t^*$ is the last time the edge $e_{ij}$ was used in exchanges
with other peers (i.e. its last ``activation") and $\beta_{ij}$ is the ``speed
of forgetting". This factor depends on the number of previous accesses to this
edge in the past. Therefore, we can define this parameter as follows:
$$\beta_{ij} = \frac{\gamma}{p^t_{ij}}$$
\noindent where $\gamma$ is a speed coefficient and $p^t_{ij}$ is the
``popularity" of $e_{ij}$ at time $t$, i.e. the number of times $e_{ij}$ has
been used in the encounters happened before $t$.

At time $t_0$, when the  device is activated, we have that $f(e, t_0) =1$,
$\forall e \in E$. Subsequently, whenever $f(e,t) \leq \phi_{min}$, where
$\phi_{min}$ is a limiting threshold value, then the edge $e$ is removed (i.e.
it is ``forgotten") from the semantic network. The  $\phi_{min}$ value is termed
as the {\em forget threshold}. 

Vertices could be also affected by the forgetting process that is acting on the
edges of the semantic network. In fact, the removal of an edge $e$ may leave a
vertex $v$ at one of $e$'s endpoints completely disconnected. In this case, the vertex $v$ is also
deleted from the semantic network. After deletion, the only way to add again an
edge or a vertex to the semantic network is to receive it during successive
encounters with other nodes, which is the process we explain next. 

\begin{figure}[ht]
  \centering
  \includegraphics[width=0.6\columnwidth]{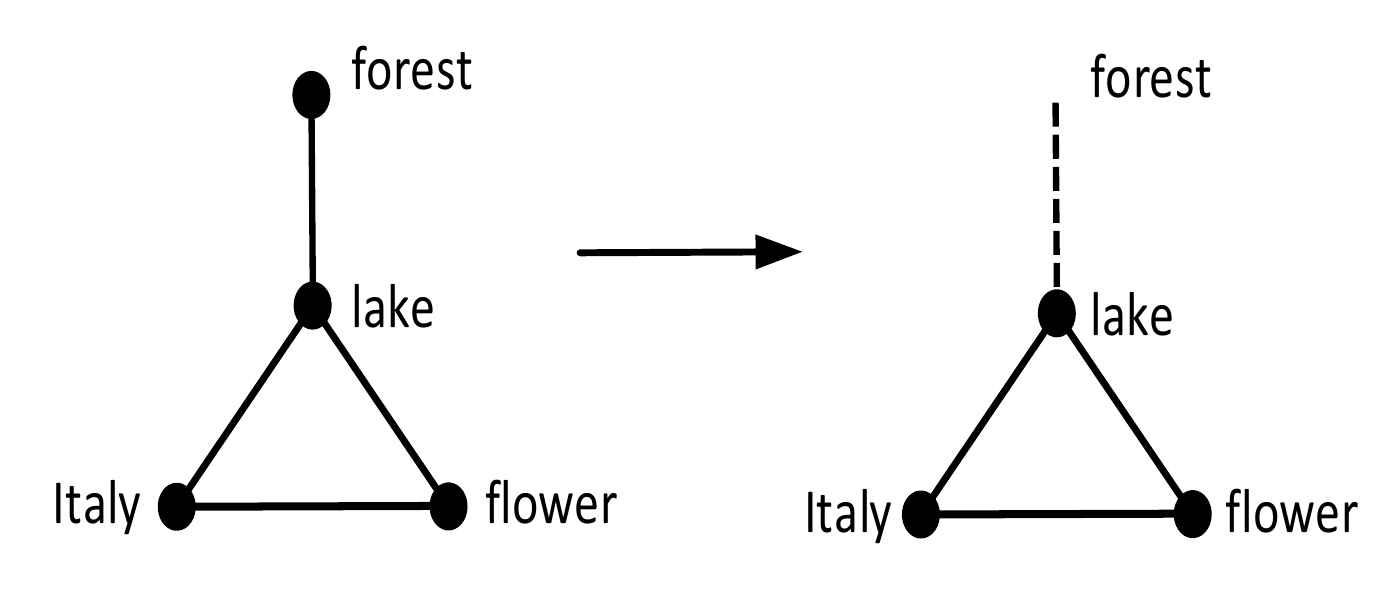}
  \caption{Effects of the forgetting process}
  \label{fig:forgVert}
\end{figure}

\subsection{Semantic knowledge dissemination}
\label{sec:fluency}
In this section, we describe how a node, when meeting another peer, is able to
retrieve from its memory the most relevant semantic information to be exchanged.
Essentially, the proposal hereafter describes the way in which concepts pass
from person to person during conversations. 
%The two nodes involved in the communication process swap roles (donor and
%recipient) in order to realise a bidirectional exchange of information. 

To model the communication process, the two nodes, involved in a bidirectional
communication process, take alternatively the ``donor" and ``recipient" role.  As
we will see in the following, the overall process used for exchanging semantic
knowledge is guided by the {\em fluency} cognitive heuristic (see
Section~\ref{ssec:fluency}). Looking at this process from a device's point of
view, in the rest of this section we make use of the following definitions. The
considered node is called the {\em donor node} and its semantic network $G=(V,E,
f(e,t))$ is the {\em donor network}. The semantic network
$G^\prime=(V^\prime,E^\prime,f^\prime(e^\prime,t))$ of the other peer ({\em
recipient node}) is termed the {\em recipient network}. The semantic information
to be exchanged is computed as a subgraph
$C=(\bar{V},\bar{E},\bar{f}(\bar{e},t))$, $C \subseteq G$, selected from the
donor network. $C$ is called the {\em contributed network}. Moreover, we assume
that resource consumption constraints limit the number of exchangeable concepts
(i.e vertices contained in $C$) to a value {\em tag\_limit}. 

We now give a detailed description of the process used by the donor node for
computing the contributed network $C$, using the fluency heuristic. The
subsequent description makes reference to the algorithmic specification of this
process given in Algorithm~\ref{alg:contnet} and Algorithm~\ref{alg:visit}.

Since the contributed network $C$ is computed as a subgraph of the donor network
$G$, we have to first define where this computation should start. Following the
analogy with a real human communication, we assume that the ``dialogue" between
donor and recipient nodes starts from a set of commonly shared concepts. This is
a set of so-called {\em key vertices} $K =\{v_k| v_k \in V \cap V^\prime\}$
(line 3 of Alg.~\ref{alg:contnet}). The inclusion of vertices in $K$ serves as a
stimulus for starting the search of information in the semantic network of the
donor node. Since in the cognitive science field the repetitive observation of a
stimulus has an influence in future searches in an associative network model, we
map this cognitive aspect into our model by increasing the popularities of all
the edges attached to a vertex each time it is included in $K$ (lines
5--7 of Alg.~\ref{alg:contnet}). 
%The relevance of a vertex is increased every time it is included in $K$ during
%information exchanges with other peers. In our system, the  {\em key vertices}
%relevance is augmented by increasing the popularities of all the edges attached
%them (lines 5--7).  The presence in other node's knowledge of this set of
%concepts represents a stimuli from the environment about the significance of
%that vertices. The reflect this, we  increase the popularities of all the edges
%attached to any key vertex (lines 5--7).  A human communciation can be well
%exemplified by the sequential search paradigm over associative networks, where,
%starting from an activated concepts, related nodes are retrieved in a
%sequential order. In our case 
Then, the computation of the contributed network can start from the set of  key
vertices. %vertices and edges are selected from the donor network by
These vertices are first ordered according to their relevance in memory. Relevance is
computed for each key vertex by summing up the weights of its incoming edges,
taken as a measure of the total importance (or relevance) of that concept in the
node's semantic network (line 8 of Alg.~\ref{alg:contnet}). Taking the key vertices (sorted by
relevance) one at a time, edges and vertices are visited and passed from the
donor network to the contributed network using Alg.~\ref{alg:visit}, that is based
on {\em fluency}. In this algorithm, the {\em fluency} heuristic is applied to
evaluate whether to follow an edge $e_{ij}$ or not. {\em Fluency} is a cognitive
decision-making strategy that favours {\em recognized} objects (i.e. objects
seen more than a given amount of times, see Section \ref{ssec:fluency}) against
unrecognized items. It assumes that the former are more relevant than the latter
ones, in the information selection task.
 %Since it discirminates among {\em recognized} objects (see Sec.2), 
Hence, we start by excluding all {\em unrecognized} edges, i.e. the edges whose
popularity is below a {\em recognition threshold} $\theta_{rec}$ (line 5 of
Alg.~\ref{alg:visit}).
\begin{algorithm}[H]
  \caption{Contributed Network computation at time $t^*$}
  \label{alg:contnet}
  \begin{algorithmic}[1]
    \State Let $G=(V,E,f(e,t))$ be the donor network;
    \State Let $C=(\bar{V},\bar{E},\bar{f}(\bar{e},t))$ be the contributed network;
    \State Let $K$ be the set of {\em key vertices}, $K \subseteq V$
    \For{ each $v_i \in K$}
      \For{ each $e_{ij} \in E$}
        \State increase popularity of $e_{ij}$
      \EndFor
      \State Let $rel_{ij} = \sum_{e_{ij} \in E} f(e_{ij},t^*)$
    \EndFor
    \For{ each $v_i \in K$ taken in desc. order w.r.t. $rel_{ij}$}
      \State $visit(v_i,1,t^*-t)$
    \EndFor
    \State Send $C$ to the other node
  \end{algorithmic}
\end{algorithm}

Acting only over recognized items, {\em fluency} makes a subsequent
discrimination based on the perceived speed of retrieval from memory. In  order
to replicate this fact in our system, we consider that the following facts
affect the ease of retrieval, and, thus, the relevance, of semantic concepts
during an interaction:

\begin{itemize}
  \item the highest the recall value of an edge (i.e. it is more
    easy to recall it), the most relevant the edge is;
  \item the relevance of an
      edge decreases as long as we get farther from a key vertex, i.e. it is
      more difficult to recall it, given the actual ``topics of discussion" (the
      key vertices);
  \item anyway, the longer the contact time between two nodes
	(i.e. the longer the discussion is), the  more time is available to
	navigate the donor network and include edges and vertices in the
	contributed network, i.e. more concepts can be recalled with longer
	``discussions".
\end{itemize}
%  andThis surely depends on the strength in memory of the association
%  represented by an edge. Anyway, the further we get from a key concept and the
%  shorter a communication, the harder becomes for the brain to retrieve an
%  association from memory. To replicate this fact,
In order to take all these observations into account, we compute, for each
outgoing  edge $e_{ij}$ of a vertex $v_{i}$, a {\em ``retrieval weight"}
quantity. This quantity is computed for an interaction that starts at a time $t$
and ends at time $t^*$. It is defined as: \begin{equation} w(e_{ij},n,t^*-t) =
  f(e_{ij},t^*) \frac{1 - e^{-\tau (t^*-t)}}{n} \end{equation} \noindent where
$f(e_{ij},t^*)$ is the memory strength value of $e_{ij}$ at time $t^*$, $n$ is
the the number of hops in the shortest path to the nearest key vertex and $\tau$
is a ``speed" factor that regulates the dependency of the weight on the
communication duration $(t^*-t)$. The retrieval weight is taken here as a
surrogate of the speed of retrieval needed by the fluency heuristic, since it
favours the edges with higher strength in memory and that are more
well-connected to the key vertices.

Given a vertex $v_i$, its outgoing edges are sorted with respect
to their retrieval
weight value (line 6). Taking them one at a time in descending order, we include
the selected edge in the contributed network and continue the donor network
exploration from this connection\footnote{This corresponds to a depth first
  visit, as discussed in Section \ref{ssec:memory}} (lines
  6--13 of Alg.~\ref{alg:visit}). All the edges
  whose retrieval value is below a threshold $\omega_{min}$ are not considered
  (line 7 of Alg.~\ref{alg:visit}). Note that the strength in memory of
  selected edges is set to the
  maximum, since inclusion in the exchanged data corresponds to an ``activation"
  in memory of those connections.

\begin{algorithm}[ht]
  \caption{Function $visit(v_i,n,t^*-t)$}
  \label{alg:visit}
  \begin{algorithmic}[1]
    \State Let $G=(V,E,f(e,t))$ be the donor network;
    \State Let $C=(\bar{V},\bar{E},\bar{f}(\bar{e},t))$ be the contributed network;
    \If{$|\bar{V}| < $ {\em tag\_limit}}
      \State $\bar{V} \cup = v_i$
      \State Let $R=\{e_{ij} \in E | p^{t^*}_{ij} \geq \theta_{rec}\}$
        \For{{\small each $e_{ij} \in R$ in desc. order w.r.t. $w(e_{ij},n,t^*-t)$}}
          \If{$w(e_{ij},n, t^*-t)\geq w_{min}$}
	    \State $\bar{E} \cup = e_{ij}$
	    \State $f(e_{ij},t^*) = 1$
	    \State $\bar{f}(\bar{e},t^*)=1$
	    \State $visit(v_j,n+1,t^*-t)$
	  \EndIf
	\EndFor
    \EndIf
  \end{algorithmic}
\end{algorithm}

 Whenever the number of vertices added to the contributed network has reached
 the limit of tags that can be exchanged during the contact, i.e.  $|\bar{V}| =$
 {\em tag\_limit}, or no other paths (i.e. edges) can be selected from the donor
 network, the contributed network computation ends and the resulting graph is
 passed to the recipient node. This peer merges the contributed network to its
 own network (i.e. the recipient network) by simply adding all the missing
 vertices and edges. This process corresponds to an enrichment of the semantic
 knowledge of the recipient peer in terms of both concepts (i.e. vertices) and
 relationships between them (i.e. edges). All the edges received from the donor
 network (either new or already present in the recipient network) set their
 weights in memory to 1, since they are ``activated" by the ``conversation".

\subsection{Semantic content dissemination}
\label{sec:tallying}
The previous selection of the most relevant semantic concepts, with respect to
the current interaction, drives the next step in the data exchange process: the
selection of relevant data items to exchange. In order to carry out this
operation, we exploit another simple decision rule derived from the cognitive
science field: the {\em tallying} heuristic. For this step, we refer to the
pseudo-code given in Alg.~\ref{alg:tallying}. 
The rationale to select data items to exchange is based on the match between
their tags and the concepts that have been exchanged between the semantic
networks. Intuitively, we exchange data items that have  as many tags as
possible within the set of concepts ``used'' in the ``discussion'' between the
nodes. The tallying heuristic models exactly this behaviour. Given a set of cues
of cardinality $m$ (which in our case is the set of tags exchanged during the
contact as explained in Section \ref{sec:fluency}), data items are ranked based
on the number of favourable cues they possess. Therefore, we rank data items for
possible exchange based of the cardinality of the intersection between the cues
and the tags associated to them (lines 5--7).
%In our scenario, each data items is associated to a set of tags, that we term
%as its semantic description ($semanticDesc(i)$ in Alg.~\ref{alg:tallying}, line
%6). We use this semantic description to relate the data items to the content of
%the actual interaction between two nodes. In fact, we consider the vertices
%selected for inclusion in the  {\em contributed} network as the $m$ ``cues"
%(out from all the $M$ vertices in the donor network) needed by the {\em
%tallying} heuristic. The relevance of data items to the actual communication
%can the be computed using the tallying heuristic. As a matter of fact, counting
%the number of favourable cues for a data item simply corresponds to counting
%the cardinality of the intersection between the data item semantic description
%and the set of vertices included in the {\em contributed} network (lines 5--7). 

We also consider that the other party sends to the node the list of the IDs of
the items it already owns. Thus, those data items can be directly pruned out
from the selection process (line 4). Moreover, like for the exchange of semantic
concepts, we assume that there is a maximum number of exchangeable data items
{\em data\_limit}. Thus, once the data items have been ranked according to {\em
tallying}, the first {\em data\_limit} ones are selected to be passed to the
other interacting peer (lines 8--9).
\begin{algorithm}[ht]
  \caption{Tallying algorithm}
  \label{alg:tallying}
  \begin{algorithmic}[1]
    \State Let $\bar{V}$ be the nodes of the contributed network;
    \State Let $I$ be the set of data items of the node
    \State Let $J$ be the data items owned by the other peer
    \State Let $I^\prime = I - J$
    \For{ each $i \in I^\prime$}
      \State Let $tall(i) = |semanticDesc(i) \cap \bar{V}|$
    \EndFor
    \State Rank $I^\prime$ in descending order according to the $tall$ values
    \State Send the first  {\em data\_limit} items of $I^\prime$ to the other node
  \end{algorithmic}
\end{algorithm}

Note that the chosen ``cues" used by our version of {\em tallying} are regarded
as relevant since they derive from a previous decision-making cognitive process
that is pertinent with the actual contact. Moreover, the number and identity of
these cues vary from one encounter to the other, reflecting the peculiarity of
each separate meeting and the changes made by previously experienced contacts in
the environment. 

\begin{figure}
  \centering
  \includegraphics[width=0.9\textwidth]{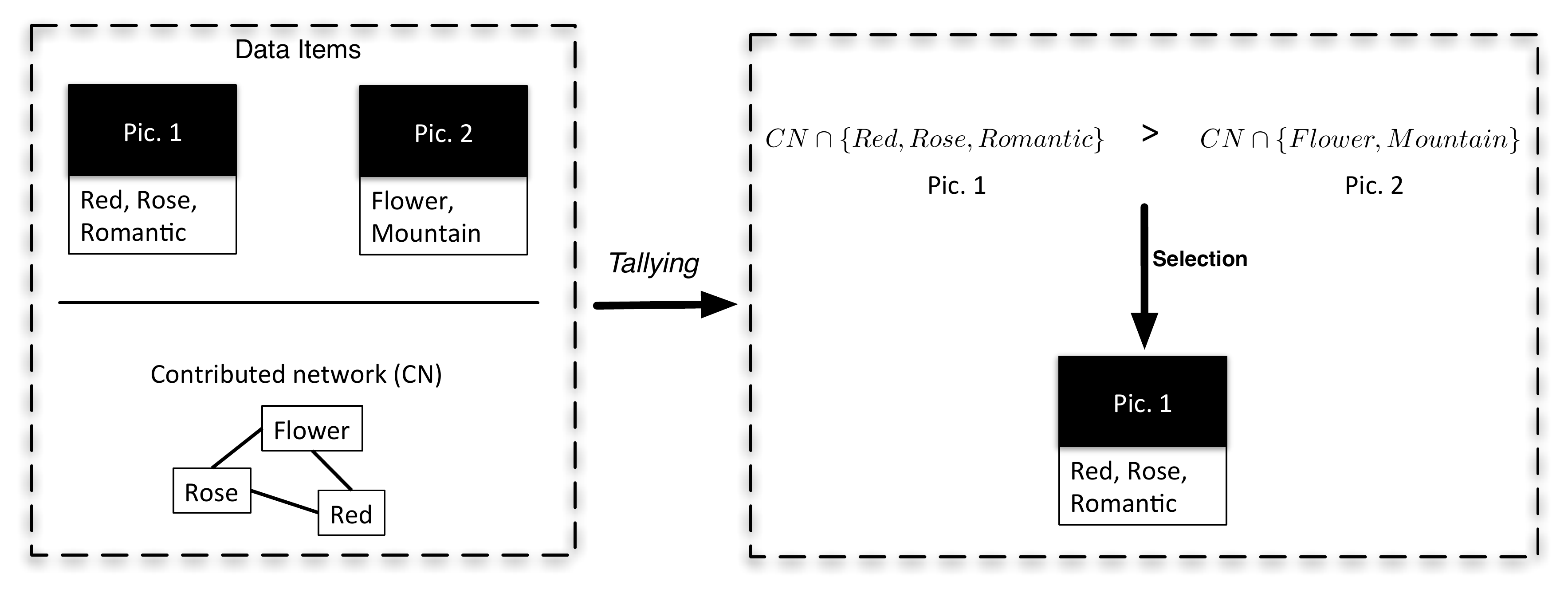}
  \caption{Example of Tallying Heuristic applied to the semantic content
  exchange process}
  \label{fig:algtallying}
\end{figure}

An example of the tallying heuristic applied to the data items selection problem
is shown in Fig.~\ref{fig:algtallying}. In this figure, we have two data items
(two pictures) along with their own semantic descriptions and an already
computed contributed network. Since the cardinality of the  intersection of the
semantic description of the first picture is greater than that of the second
picture (2 vs. 1), the first picture is selected for being exchanged.

\section{Experimental results}
\label{sec:expRes}
In this section, we analyse the performance of our solution through a set of
experiments performed in synthetic scenarios. Precisely, we test our approach
against an alternative  solution, comparing both approaches  in terms
of knowledge and content dissemination effectiveness, whereby, by knowledge
dissemination we mean how concepts spread from node's to node's semantic
network. We analyse and compare the structural properties of the nodes'
semantic networks built according to the two approaches and, finally, we provide
a sensitivity analysis of our approach.

%:-- Simulated Environment
\subsection{Simulation settings}
%
%HCMM
\subsubsection{Synthetic mobility scenarios}
Mobility traces are generated by
HCMM~\cite{Boldrini10}, a well known mobility model already used in several
papers in the literature to evaluate data forwarding and data dissemination
algorithms for opportunistic networks. HCMM incorporates temporal, social and
spacial notions in order to obtain a proper representation of  real users
movements~\cite{Karamshuk:2011qy}.  In HCMM the simulation area is
organised as a rectangular grid, where a single grid cell represents the
physical location of a community of nodes. In each community two kinds of mobile
nodes  are allowed:  \emph{travellers} and \emph{non-travellers}. Non-travellers
roam only inside their community, while travellers, according to a given
probability distribution, from time to time visit other social
communities different from the one they belong to. When moving inside a
community, for each simulated time step, each node (both travellers and
non-travellers) randomly select a velocity and a new
position to reach inside the community area. Velocity and position are selected
according to uniform probabilities (for velocity, inside a given velocity
range). In this context, the only way
to exchange information is by means of nodes mobility, and travellers play an
important role because they are the unique bridge between communities
(besides possible border effects between nodes of adjacent communities).

In this paper we considered three different mobility scenarios. Precisely, in
\emph{Scenario 1} there are $99$ mobile nodes moving in a $1km^2$ area and
grouped in a single social community. In \emph{Scenario 2} we consider a less
crowded configuration. In the same simulation area ($1000m \times 1000m$) as for Scenario
1, we have only $50$ mobile nodes roaming in a single social community. Finally,
in \emph{Scenario 3}, $99$ nodes are divided in three physically separated
groups ($33$ nodes for each group) representing three different social
communities.  Nodes move in an area of $1km^2$ divided in a $6\times6$ grid,
and communities are placed far from each other so to avoid any border
effect. 
% Nodes move in an area of $1000m^2$ divided in a $6 \times 6$ grid, where a
% single grid cell represents the physical location of a community. In this
% specific scenario, communities are placed far from each other so to avoid any
% border effect, e.g., involuntary communication between groups. Moreover, in
% each community there are two kind of nodes: \emph{travellers} and
% \emph{non-travellers}. Non-travellers roam only inside their community, while
% travellers, from time to time, use to visit other social communities different
% from the one they belong to. In this context, the only way to exchange
% information is by mean of nodes mobility, and travellers play an important
% role because they are the unique bridge between communities.
Details on  configuration parameters for the scenarios
can be found in Table~\ref{tab:confHCMM}.
 
 \begin{table}[ht]
   \begin{center}
     \caption{Detailed scenario configurations\label{tab:confHCMM}}{
       \begin{tabular}{|c|c|}
	 \hline
	 \textbf{Parameter} & \textbf{Value}\\
	 \hline
	 Node speed & Uniform in $[1,1.86m/s]$\\
	 Transmission range & $20m$\\
	 Simulation area & $1000\times 1000m$\\
	 Number of cells & $1\times 1, 6\times 6$\\
	 Number of nodes & $99$, $50$\\
	 Number of communities & $1$, $3$\\
	 Number of travellers & $0$, $6$ ($2$ per group)\\
	 Simulation time & $25000s$\\
	 \hline
       \end{tabular}
     }
   \end{center}
\end{table}
	
%DATASET
\subsubsection{Dataset description}
\begin{figure}[ht] \centering \subfloat[]{
	    \includegraphics[width=.5\textwidth]{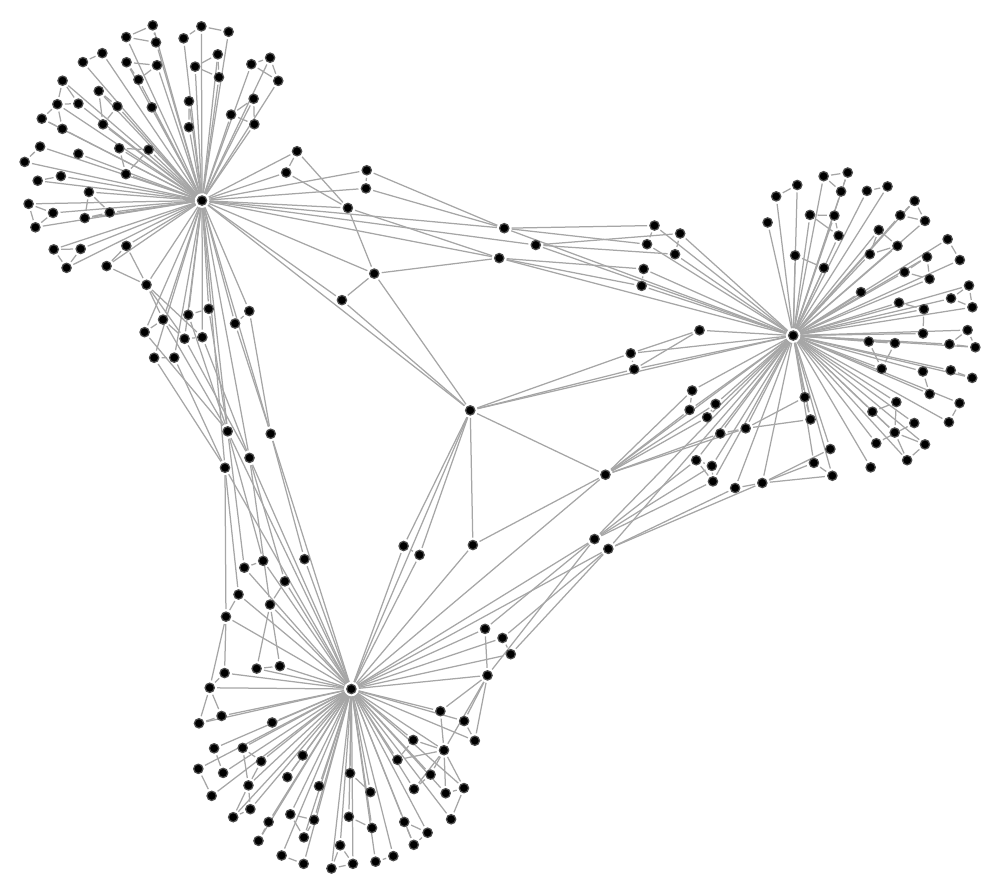}
	    \label{fig:beginGraphD1} } \subfloat[]{
	      \includegraphics[width=.5\textwidth]{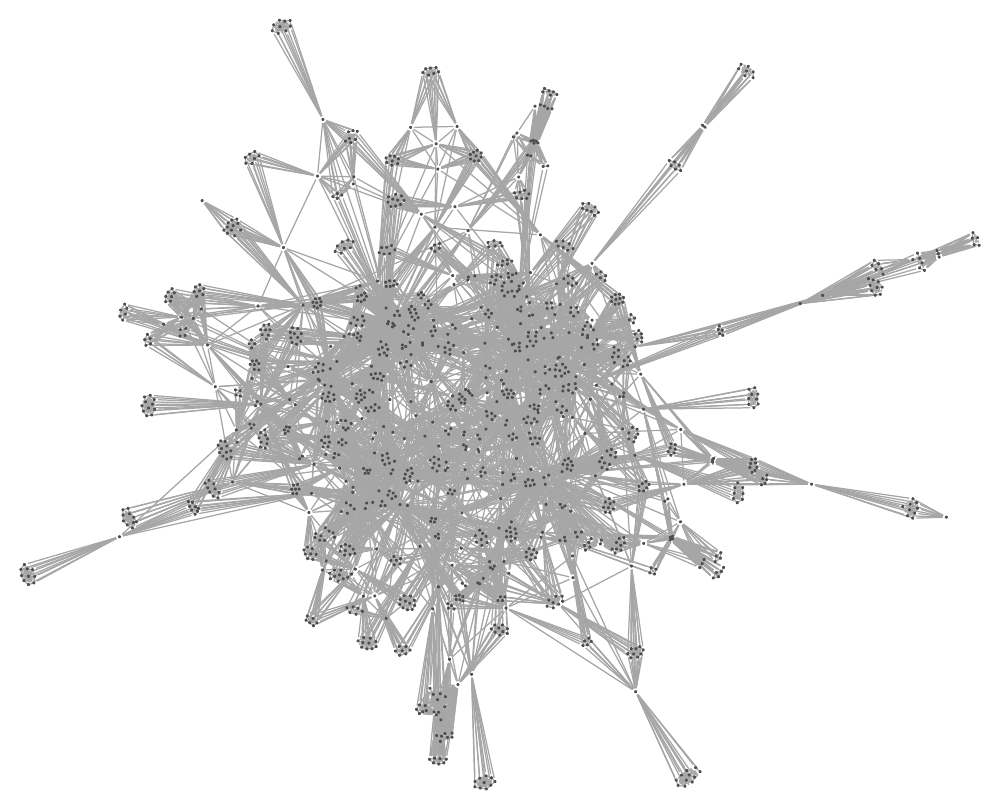}
	      \label{fig:beginGraphD2} } \caption{Initial knowledge graphs
	      (namely $G1$ (a) and $G2$ (b))  defined as the union of all nodes'
	    semantic networks at the beginning of the simulation for dataset
	  $D1$ (a) and $D2$ (b)} \label{fig:initialKnowledge}
	%\vspace{-1.2em}
	\end{figure}

Data assigned to the
nodes is selected from
the CoPhIR dataset~\cite{CoPhIR}. This  dataset is made up of more than $100$M
images coming from Flickr. For each user's image, the list of associated tags is
available. To test our solution, we created two datasets, $D1$ and $D2$,
according to the following procedure. 
	%descrizione dataset D1
	In $D1$,  images were selected in order to have the initial users'
	semantic knowledge strongly clustered around three main concepts poorly
	connected to each other. Specifically, a main concept is a hashtag
	containing a very common word representing a very general category of
	pictures, e.g. ``mountain", ``sea", ``lake". In this case, selected images
	have from $2$ to $4$ tags and only one of the three main concepts.
	Figure~\ref{fig:beginGraphD1} represents the entire knowledge present in
	the network at the beginning of each simulation, i.e. the graph of the
	union of all nodes'  initial semantic networks. The resulting graph $G1$
	is made by 221 vertexes and 455 edges.  This dataset represents a sort
	of a controlled environment we used to perform a first evaluation on the
	performance of our approach, while the rest of investigations have been
	made on the more complex dataset $D2$, hereafter described.
	
		%descrizione dataset D2
$D2$ represents a more real situation in which the resulting graph built from
the union of all nodes' initial semantic networks has a more complex structure.
Images in $D2$ have from $10$ to $15$ tags each and no other constraints to
drive the images selection have been considered (as we did for $D1$).  Figure
\ref{fig:beginGraphD2} shows the entire knowledge present in the network at the
beginning of each simulation. The resulting graph $G2$ is made by 1302 vertexes
and $8845$ edges. In Table \ref{tab:graphProps} more details about both $G1$ and
$G2$ are reported.  We pointed our attention on such initial configurations in
order to both study the ability of each user to retrieve  the information
semantically related to its initial interests and to analyze the overall
permeation of data in the network.  Note that the interests of the users are
automatically defined by the tags in their local semantic networks. Initially,
they are thus defined by the tags of the locally available pictures, and evolve
over time based on the diffusion of tags due to the algorithms presented in
Section \ref{sec:approccio}.
%It is worth noting that each node in the simulation represents a real Flickr
%user with its own tagged pictures. 
	 \begin{table} \begin{center} \caption{Properties of graphs
	      build on datasets $D1$ and $D2$. \label{tab:graphProps}}{
		\begin{tabular}{|c|c|c|} \hline \textbf{Property} & \textbf{G1}
		  & \textbf{G2}\\ \hline N. of vertexes & $221$ & $1302$\\ N. of
		  edges & $455$ & $8845$\\ Diameter & $4$ & $8$\\
%Radius & $2$ & $5$\\
\hline \end{tabular} } \end{center} \end{table}

% MISURE	
\subsubsection{Performance evaluation indexes}
We observed and studied the evolution
of the knowledge and content acquisition processes generated by the interactions
between users.  We defined two indexes to measure the knowledge and
content dissemination performance, respectively. Moreover, we used the
F-measure~\cite{Rijsbergen:1979nr}, a commonly used metric in the field of
\emph{information retrieval}, to evaluate the relation -- in terms of accuracy
-- between the knowledge acquired and the contents retrieved by nodes, during
the simulation (a more formal definition of the F-measure is provided in
Equation (\ref{eq:f-measure})). In Table \ref{tab:notation} we
provide the notation we use to define the above mentioned indexes.

\begin{table}[ht!]
  \begin{center}
    \caption{List of notations and definitions used in this paper.\label{tab:notation}}
    {\begin{tabular}{|c|c|}
      \hline
      \textbf{Symbol} & \textbf{Meaning}\\
      \hline
      $N$ & Set of nodes in the scenario\\
      $K_n$ & Semantic network of the node $n$\\
      $V_{n}$ & Set of vertexes (tags) in the Semantic Network $K$ of node $n$ \\
      %$v_{ij}$ & the $j$-th tags in $V_{SN_i}$ of node $i$\\
      $C_n$ & Set of contents owned by the node $n$\\
      $T_{c}$ & Set of tags of the content $c \in C_n$ for the node $n$ \\
      $G=\bigcup_{\forall n \in N}K_n$ & Graph of complete knowledge (e.g. $G1$ or $G2$)\\
      $V_G$ & Set of vertexes of graph $G$\\
      %$E_G$ & Set of edges of graph $G$\\
      $D$ & Set of all contents in the system\\
      $D_v$ & Set of all contents in the system tagged with the tag $v\in V_G$\\
      \hline
    \end{tabular}}
  \end{center}
\end{table}
	 
We measure the Knowledge Dissemination (KD) by computing
how much of the starting global  knowledge reaches nodes at the end of the
simulation. The KD index is defined in Equation
(\ref{eq:kd}).

\begin{equation}
\label{eq:kd}
\overline{KD} = \frac{1}{|N|}\sum_{n \in N} \frac{|V_n|}{|V_{G}|}
\end{equation}

$KD$ in Equation (\ref{eq:kd}) measures the average (over all nodes) percentage of
the entire knowledge that is available at nodes at the end of the simulations.
Reaching a $KD$ in the range of $100\%$ would mean that all semantic data
reach all nodes. Given the structure   of the initial overall semantic network
and the knowledge acquisition process we don't expect to reach $100\%$ for this
index. Precisely, in our system, users acquire only the information semantically
correlated to their knowledge, while throwing out less relevant concepts. This
triggers a filtering process that generally prevents users from collecting all
the existing concepts in the environment.  
	 
We also define a second measure, the \emph{coverage}, as the fraction of items
owned by a node over all the items that contain a tag matching one of the
concepts in its semantic network This index is defined in
Equation (\ref{eq:coverage}). Precisely,
for a given node $n$ and a given tag $v$ in its SN we calculate the ratio
between  the number of contents owned by the node having $v$ in their semantic
description, over the number of all contents in the system whose semantic
description contains $v$ (part B of (\ref{eq:coverage})). Then we average over
all the tags in the node's SN and over all nodes in the system (part A of
(\ref{eq:coverage})).

\begin{equation}
  \label{eq:coverage}
  \overline{CVG} =
    \overbrace{\frac{1}{|N|} \sum_{\forall n \in N} \frac{1}{|V_n|}}^A
    \overbrace{\sum_{\forall v \in V_n}\frac{1}{|D_v|}\sum_{\forall c \in C_n}
      \mathbbm{1}_{ v \in T_c}}^B
\end{equation}

For this index an ideal dissemination system would reach $100\%$,
as this means that users receive all the data items related to concepts they
have in their semantic network.

Finally, in order to evaluate the accuracy of the semantic information acquired
with respect to the contents retrieved, we used the
\emph{F-measure}. It is made of two
indexes called \emph{precision} and \emph{recall} denoted by $p$ and $r$,
respectively.  In our setup the \emph{precision} index (between 0 and 1) measures how appropriate
are the contents retrieved by a node with respect to its
semantic network.  For a given node $n$ the precision is defined
as follows:

\begin{equation}
  p_n =
  \frac{|\bigcup_{\forall c\in C_n}T_c \cap V_n|}
    {|\bigcup_{\forall c\in C_n}T_c|}
\end{equation}
      
The \emph{recall} index (between 0 and 1) measures how appropriate is the node's semantic network
with respect to the contents it has collected during the
simulation. For a given node
$n$ the \emph{recall} is defined as follows:

\begin{equation}
  r_n = \frac{|\bigcup_{\forall c\in C_n}T_c \cap V_n|}{|V_n|}
\end{equation}
      
In our context, precision equal to $1$ means that there are no tags associated
to content on node $n$ that are not in its semantic network (or, in other words,
that all retrieved content are associated to tags ``known'' by the node). Recall
equal to $1$ means that for each tag in the semantic network there is at least
one retrieved content associated to that tag. The
\emph{F-measure} for a single node $n$ is defined as:

\begin{equation}
  \label{eq:f-measure}
  F_n = 2 \frac{p_n * r_n}{p_n+r_n}
\end{equation}

Finally, we considered the average \emph{F-measure} over all nodes:

\begin{equation}
  \overline{F} = \frac{1}{|N|}\sum_{\forall n \in N}F_n
\end{equation}

Note that $F_n$ is between $0$ and $1$, and if it is higher the more recall and
precision are both high and close to each other. Therefore it measures how good
a retrieval system is from both the precision and recall standpoints.

%COMPETITOR
\subsubsection{Benchmark algorithm description}

To best of our knowledge, there are no other solutions in literature taking into
account the problem of  disseminating both semantic knowledge and contents in
Opportunistic Networks. Therefore, we defined an alternative dissemination
algorithm that shares the general characteristics of our approach -- the
semantic knowledge is organised in semantic networks, and nodes exchange both
semantic information and contents -- but the rules according to which nodes
exchange semantic information and contents are driven by a pure random process. 

Regarding the exchange of semantic knowledge, as in the cognitive approach,
nodes populate the respective \emph{contributed networks} from a vertex they
have in common but, differently from the cognitive approach, they continue
appending semantic concepts to the contributed network by navigating their
semantic network according to a random walk. Namely, the next vertex to be
appended to the contributed network is selected with probability $1/k$ where $k$
is the out degree of the current vertex of the node's semantic network.  The
selection of the contents to be exchanged with the other peer is random, as
well. Each node uniformly selects \emph{data\_limit} contents choosing between
those having at least one tag (in their semantic description) in common with the
contributed network. 

By comparing our cognitive approach with this benchmark,
we aim at investigating
the  impact on the dissemination process (both for semantic information and
data items) deriving from the use of a structured and well defined information
selection process driven by cognitive models (in particular, cognitive
heuristics) against one where semantic information is also organised in
  an associative network, but the choice of the information and contents to
exchange is not driven by cognitive models. The selection policy used by the
benchmark is the simplest one that could be devised. Note, however, that such
uniform random policies may perform quite well in homogeneous mobility cases,
such as Scenarios 1 and 2 in our case. For example, this has been found to be
the best policy for topic-based data dissemination in opportunistic networks in homogeneous
mobility settings~\cite{Lenders08,Chiara10}.
	% As explained in Sec.\ref{sec:approccio}, nodes start to share their
	% semantic networks and the corresponding contents upon contacts.
 
Results reported in the following  simulations are the statistics collected on
10 different tests obtained from 10 different mobility traces of the HCMM model
and averaged across all nodes. Each experiment is a transient simulation ran for
$25000$ sec.  In the rest of this section, we use some conventions in order to
make the values of the \emph{forget} and \emph{retrieval thresholds} more
intuitive for the reader. Remember that the {\em forget} threshold (see
Section~\ref{sec:forget}) defines a weight value under which edges are dropped
from a semantic network, while the {\em retrieval weight} threshold (see
Section~\ref{sec:fluency}) sets the limiting weight under which edges are not
considered for inclusion in a donor network during an exchange. In the
following, we define these two parameters with respect to the time passed since
the last usage of an edge in an exchange. The longer this time, the lower the
relevance of that edge in a node's memory. Therefore, it is more difficult to
remember that edge (i.e., include it in exchanges) and it becomes more likely to
forget about it. In the rest of this section, a notation like $f_{min} = 50s$
means that the \emph{forget} threshold $\phi_{min}$ is set in such a way that
edges with popularity $1$ are dropped from a SN in case they are not seen before
$50s$ from the last time they were used in an exchange. Using this notation, we
have that the higher $f_{min}$, the longer an edge is retained in the SN. 
%% MATTEO TODO da sistemare con commmento di andrea: ��� un pochino contorto. Si riuscirebbe mica a semplificare? Sarebbe anche utile per entrambi dire cosa succede al variare dei threshold. Fmin pi��� alto vuol dire che si dimentica prima o dopo? Wmin pi��� alto vuol dire che si include pi��� o meno?
On the other hand, a notation like $W_{min} = 25s$ means that the
\emph{retrieval value} threshold $\omega_{min}$ is computed taking into account,
as a reference case, an interaction between nodes of 2s, that allows
nodes to include
(warm up) at least edges at distance $1$ from a key concept if they are not used
(i.e. they were subject to the forget process) from no more than $25s$. The
higher $W_{min}$, the higher the number of edges that are ``warmed up", and, as
a consequence, the higher the number of vertices that could be included in a
contributed network.

\subsection{Overview on the key findings}

For the sake of presentation clarity, in this section we summarise and
anticipate the key findings we draw from our experimental results, providing,
for each one, the pointer to the section where it is presented and explained in
more detail.  We found that using cognitive-based algorithms to represent
  dynamically varying users' interests, that in turn drives the dissemination of
  available data, has multiple advantages over a benchmark solution that is not
based on human cognitive models. Specifically:

\begin{itemize}
  \item
    nodes in the network are able to recognise and select the most
      relevant information available in the
      environment; moreover the 
      representation of the semantic information collected from the environment
      is very stable , i.e. the most relevant information
      last in the nodes' memory for long time, thanks for reinforcement
      from frequent accesses. See
      Section~\ref{sssec:perfeval};
  \item
    the Knowledge Dissemination triggered by our cognitive approach
      drives a more efficient content dissemination with respect
      to the benchmark.
      See Section~\ref{sssec:coverage};
  \item
    the internal structure of the local semantic information collected by
      nodes is representative of the entire semantic information present in the
      environment, i.e. the graph properties of the semantic information
      representation in nodes' internal memory  are very close to the properties
      of the complete semantic network graph. See
      Section~\ref{ssec:graphAnalysis};
  \item
    the cognitive mechanism of information selection and representation
      combined with the corresponding data dissemination algorithm proves to be efficient in
      terms of memory occupation. In fact, in Section~\ref{ssec:sens} we show
      that even increasing the resources involved in the dissemination process
      (memory for tags and data items), we obtain only a marginal increase of
      performance, meaning that limiting memory resources does not
      drastically degrade the overall
      performance of the system.
\end{itemize}
%
%:-- Experimental results \subsection{Simulation results}
\subsection{Performance analysis of information-dissemination algorithms}
\label{sssec:perfeval}

In this section we provide a comparison of the
proposed solution against the benchmark  approach , in all scenarios
previously described. For the sake of simplicity, from now on the acronyms CA
and BA we will refer to the cognitive approach and the benchmark approach,
respectively.   

We compare the two approaches using the  Knowledge
Dissemination (KD), F-Measure and Coverage metrics. All the results reported in
this section are obtained by varying the $f_{min}$ threshold between 150 and 300
sec. For all the other parameters, we use the values reported in
Table~\ref{tab:confSceneries}.  Unless otherwise stated, the $x$ axis is plotted
in log-scale. 

\begin{table}[ht]
  \begin{center}
    \caption{Detailed scenario configurations\label{tab:confSceneries}}{%
      \begin{tabular}{|c|c|}
	\hline
	\textbf{Parameter} & \textbf{Value}\\
	\hline
	$W_{min}$ & $35s$\\
	{\em tag\_limit} & 25\\
	{\em data\_limit}  & 10\\
	\hline
      \end{tabular}
    }
  \end{center}
\end{table}

\subsubsection{Scenario 1}

\begin{figure}[htbp]
  \centering
  \subfloat[\label{fig:comp150KD3}]
  {\includegraphics[width=0.495\columnwidth]{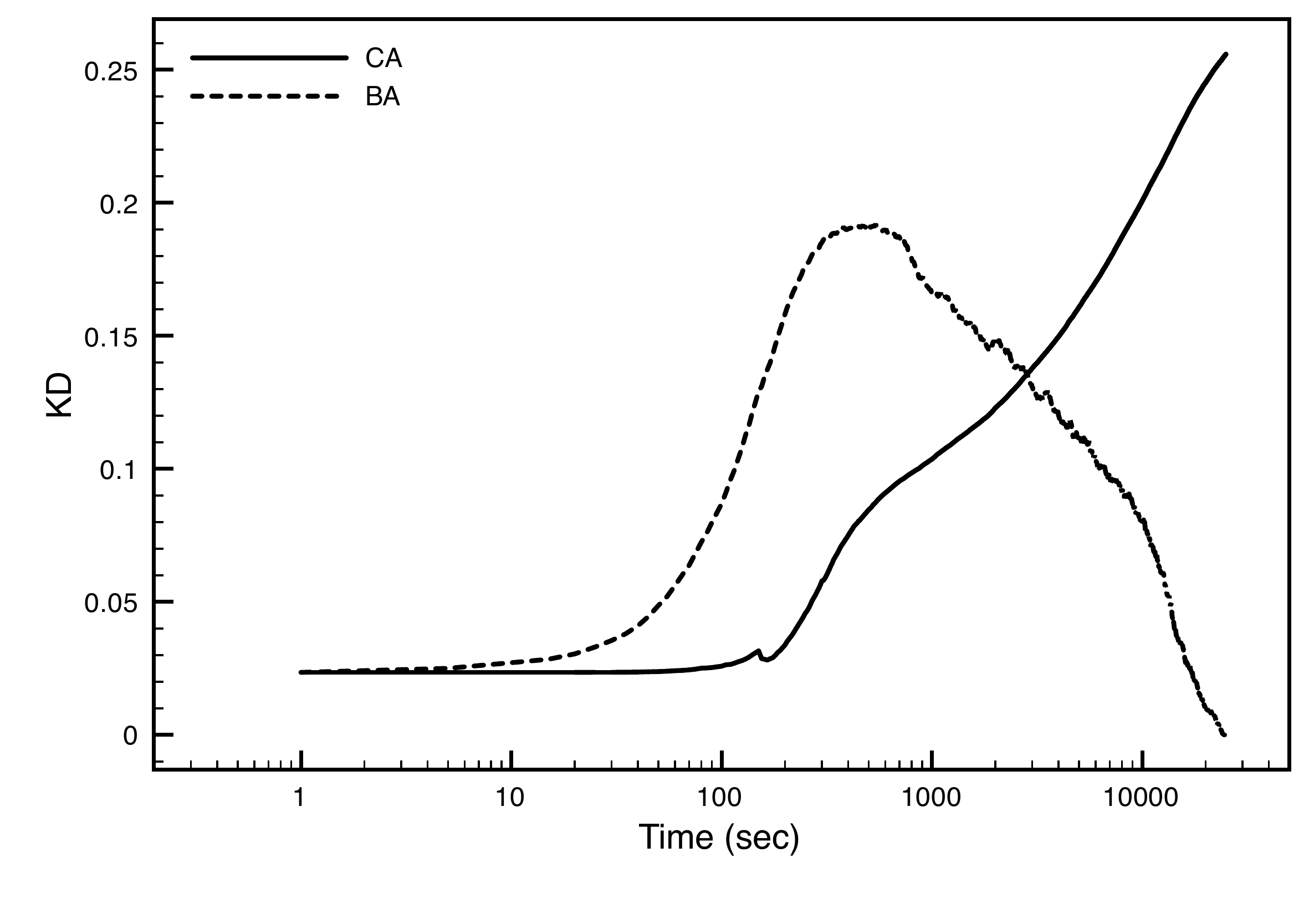}}
%  \qquad\qquad 
  \subfloat[\label{fig:comp150DE3}]
  {\includegraphics[width=0.495\columnwidth]{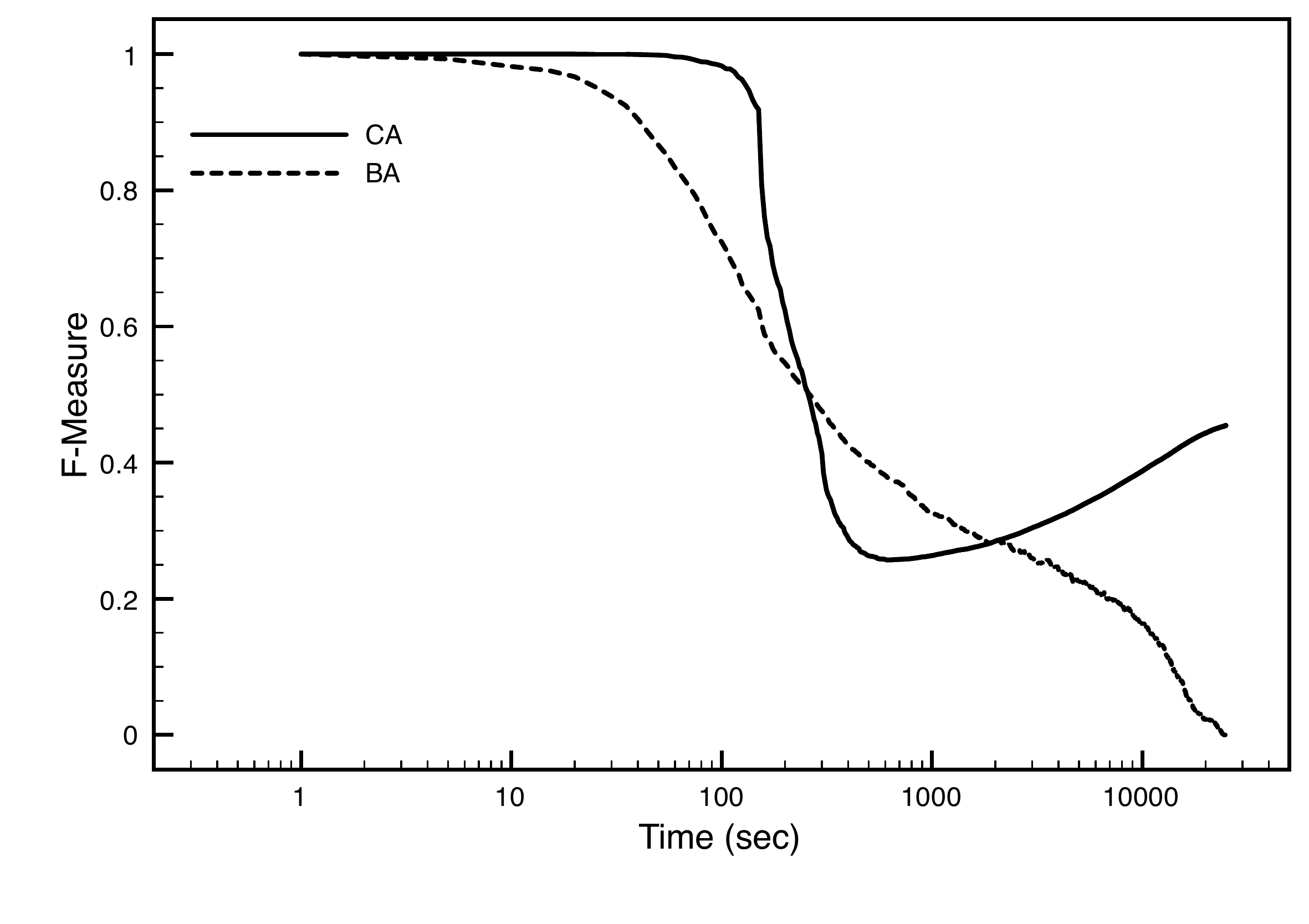}}
  \caption{KD(a) and F-measure(b) comparison with $f_{min}=150s$ on dataset $D1$
  (Scenario 1).
  \label{fig:comp150}}
\end{figure}

\begin{figure}[htbp] \centering
  \subfloat[\label{fig:comp300KD3}]
  {\includegraphics[width=0.495\columnwidth]{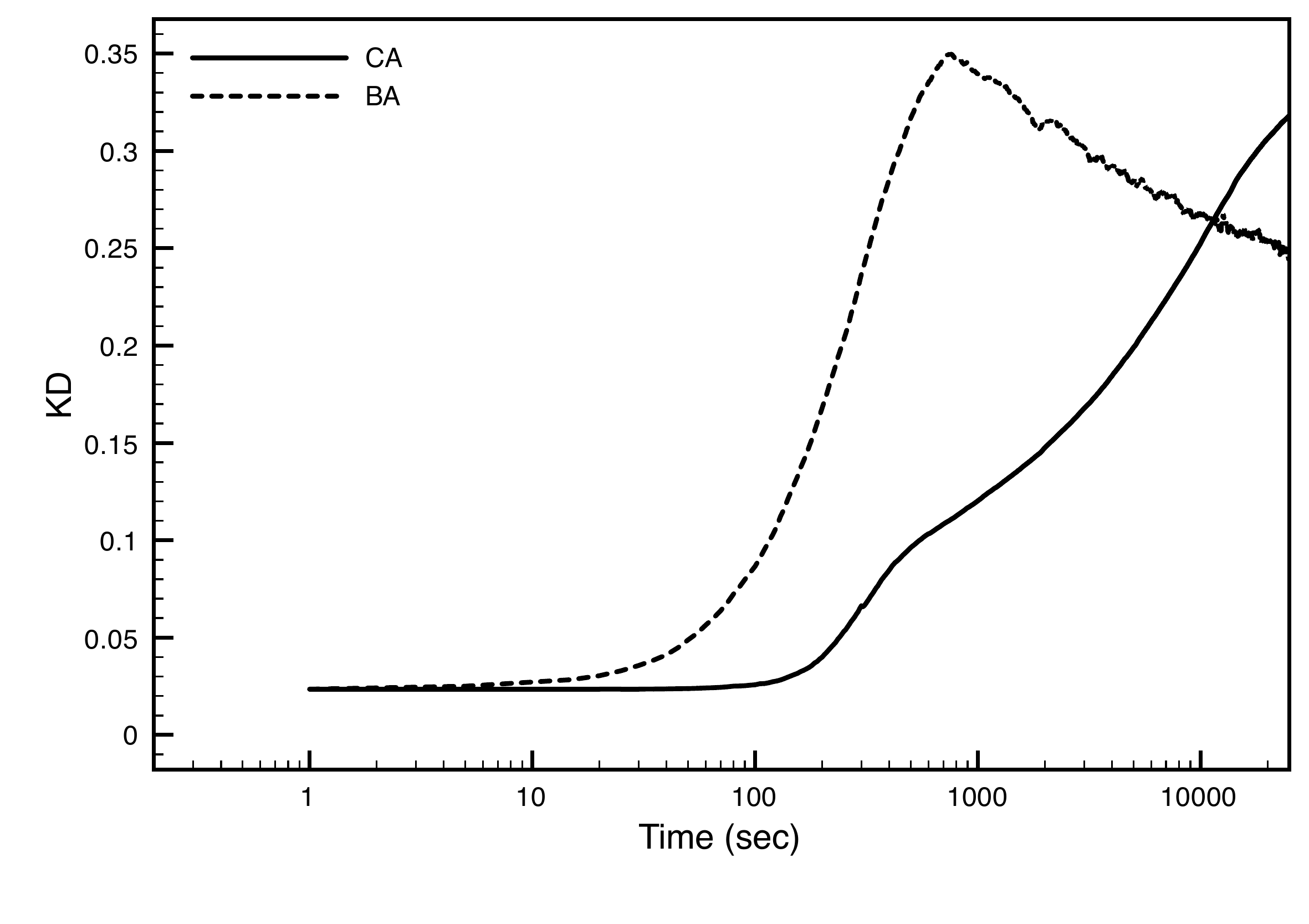}}
  \subfloat[\label{fig:comp300DE3}]
  {\includegraphics[width=0.495\columnwidth]{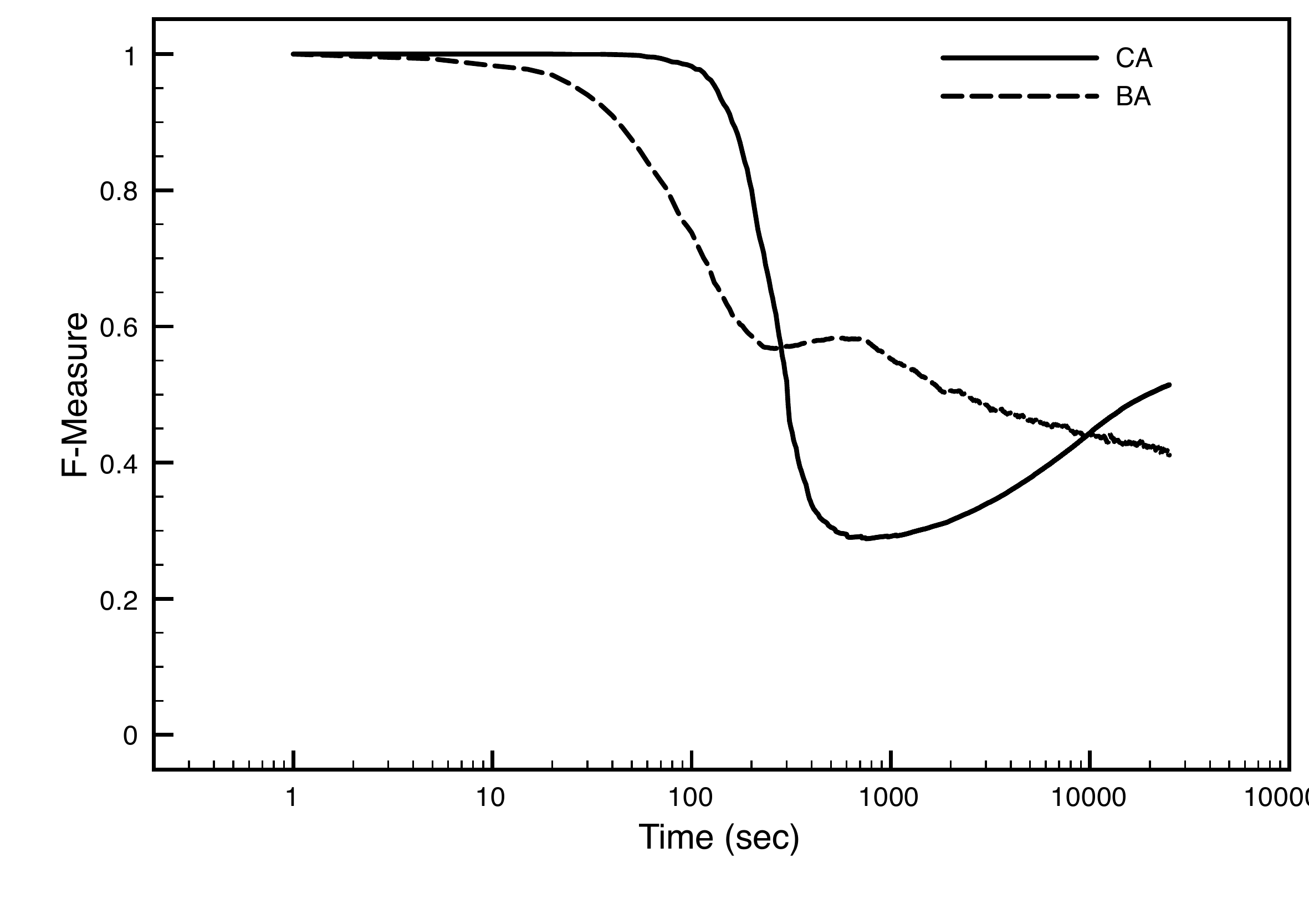}}
  \caption{KD(a) and F-measure(b) comparison with $f_{min}=300s$ on dataset $D1$
  (Scenario 1).
  \label{fig:comp300}}
\end{figure}

\begin{figure}[htbp]
  \centering
  \subfloat[$f_{min}=150s$\label{fig:comp150EW}]
  {\includegraphics[width=0.495\columnwidth]{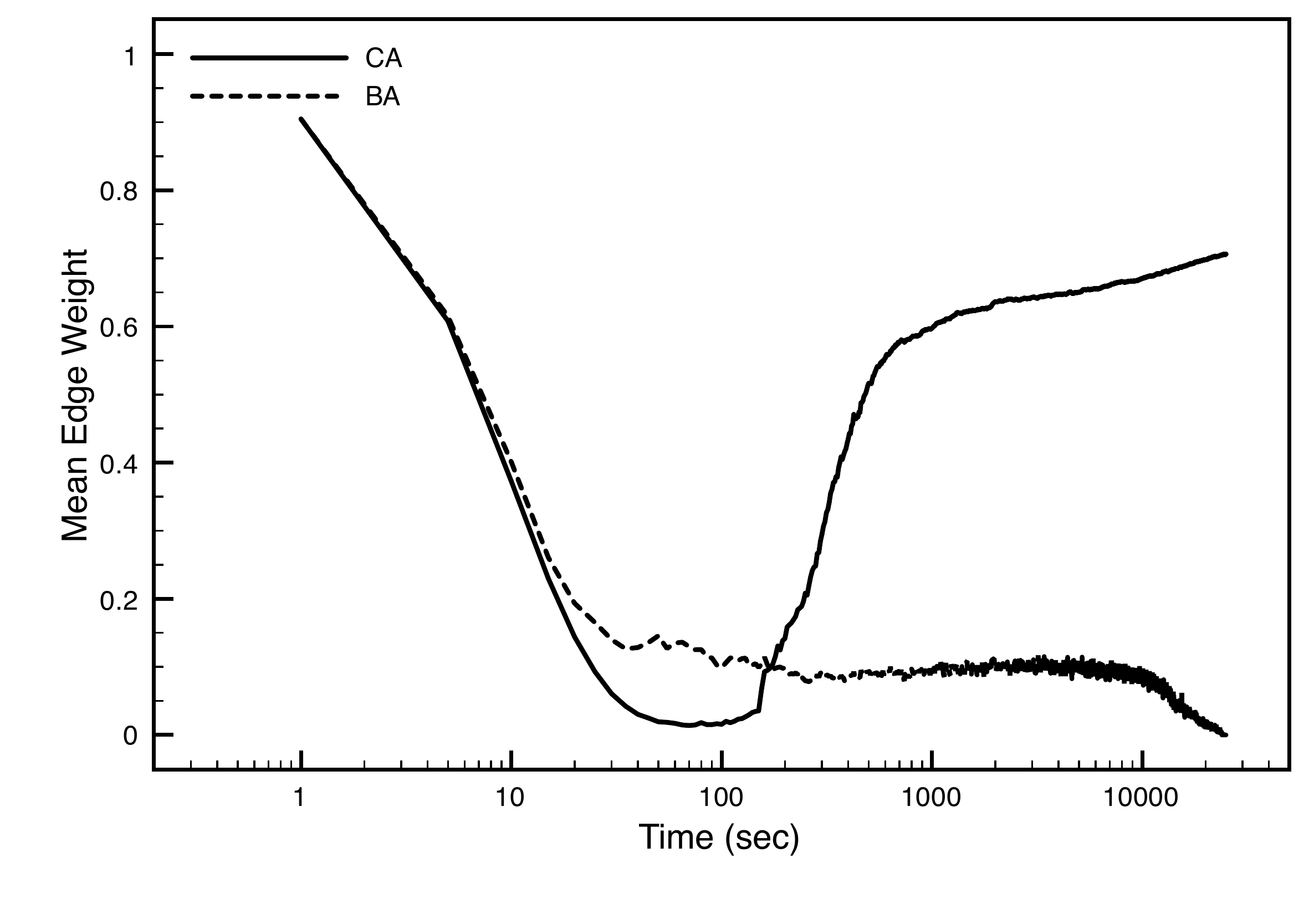}}
  \subfloat[$f_{min}=300s$\label{fig:comp300EW}]
  {\includegraphics[width=0.495\columnwidth]{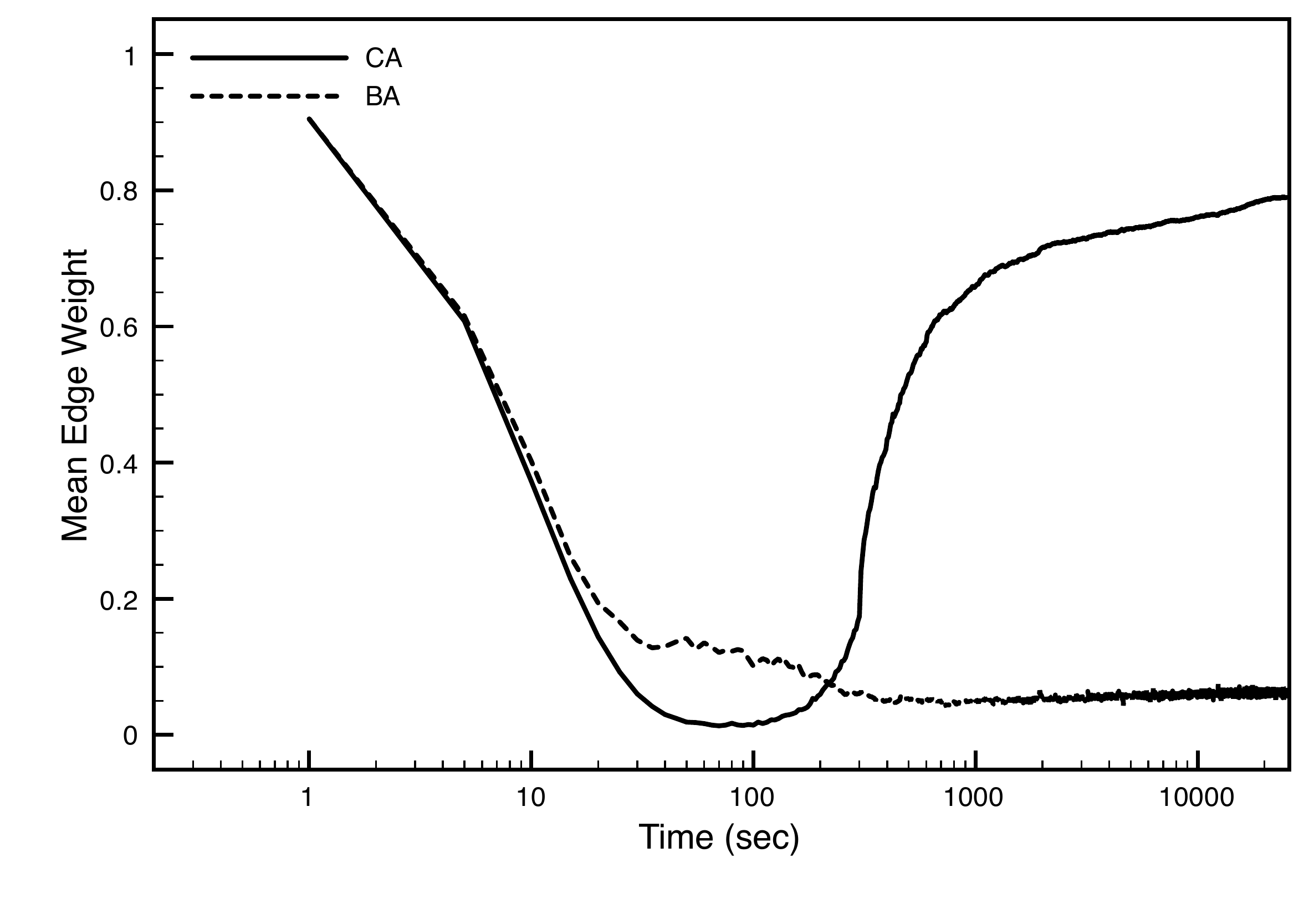}}
  \caption{Weights comparison between CA(a) and BA(b). 
  \label{fig:compEW}}
\end{figure}

The tests reported in Figures~\ref{fig:comp150}--\ref{fig:compEW} are obtained
using {\em D1}. Fig.~\ref{fig:comp150} shows the results obtained with $f_{min}
= 150s$, and in Fig.~\ref{fig:comp300} plots the results
obtained with
$f_{min}=300s$. In both these figures,  the left side shows the variation over
time of  Knowledge Dissemination, while the right
side depicts the variation of the F-measure associated to the data dissemination
process.

In both the cases presented in the figures, we can note that the F-measure of
the CA approach initially decreases, reaching a minimum, and successively starts
to increase. On the contrary, the same metric always decreases for the BA
solution. This is due to the fact that   there is some kind of inertia
in the spreading of data items with respect to the diffusion of tags. In fact,
semantic knowledge (i.e. tags) is spread first, while data items are ``pulled''
toward interested nodes as a side effect. For the CA approach, there is a
transient moment  where nodes start to increase their knowledge faster than they
exchange correlated data items, leading to a decrease in the F-measure value.
Initially, this consideration holds also for the BA approach. However, as a
consequence of the forgetting process, BA soon starts to loose relevant semantic
information, as can be seen by the KD values (discussed later in this section).
As a result, the F-measure continues to decrease at a speed that is
proportional to the deletion of semantic concepts. 

Looking at the KD metric, it is possible to note that for BA it initially shows
a rapid growth, compared to that of CA. However, as we already observe, when the
simulation proceeds, BA is not able to sustain this growth. On the contrary, it
enters a sort of ``forgetting" phase, where the semantic information gradually
gets lost. This phase is reflected also in the quality of the data dissemination
process. With the lowest $f_{min}$ value, the KD metric for the
BA solution
reaches 0, meaning that all the information has been deleted from all the
semantic networks in all the runs of the experiment.

%\begin{figure}[ht!] \centering
%\includegraphics[width=20em]{img/definitive/compKD150-dataset1}
%\caption{Comparison 150} \label{fig:comp150} \end{figure} 

The divergence between the behaviour of the two systems can be explained by the
differences in the contributed network computation mechanisms.  BA does not give
any particular preference to the edges used to explore the donor network when
computing the contributed network. In this way, it is easier to have a greater
variety in the content of contributed networks exchanged over time. As a
consequence, at the beginning the KD experiences a steep
increase.  On the
other hand, the cognitive-based solution proceeds more slowly. First of all, an
edge becomes eligible to be added to the contributed network only when {\em
recognized}. %First of all, in order to be able an edge in the contributed
%network computation, this edge must be first {\em recognised}.
Moreover,
among recognized edges, the ones that connect semantic concepts that are closer
to a {\em key vertex} are preferred to the ones that connect more distant
concepts. This mechanism leads to the creation of a sort of ``core" of semantic
information (vertices and edges) inside each semantic network. This ``core" is
increasingly reinforced over time and allows other, new concepts to be gradually
included in it, giving a more clear structure to the semantic network.

In front of these two different approaches for growing the semantic networks,
the underlying forgetting process is more likely to have detrimental effects on
the semantic networks of BA, rather than on those of CA. In fact, in the former
case there is no particular mechanism for selecting the path to follow in the donor
network, and the dimension of the semantic network increases rapidly. As a
consequence, a relevant number of edges could remain not visited for a long
time, leaving space for the forgetting process to delete them. As a further
consequence, the vertices attached to them could also be eligible for removal,
thus leading to a loss of information. 
%grows so rapidly and without taking any particular care  over a certain
%dimension, a semantic network that has grown with the random mechanism

This situation can be better viewed by analysing the variations over time of the
mean weight of the edges in all the semantic networks of the system.
Fig.~\ref{fig:compEW} presents this data for both $f_{min} = 150$ and $f_{min} =
300$. Initially, the mean edge weight starts to decrease for both BA and CA.
This is due to the fact that edges need to start being exchanged in order to
increment their weights. To this end, in CA edges must also become recognised.
After having reached a minimum, the mean weight of the edges in the
cognitive-based solution starts to rapidly increase. This is an indication of
the reinforcement of the information (and, thus, the weights of the edges) in
the semantic networks around a ``stable core". The increment in the mean edge
weight means that edges cannot generally be deleted by the forget process, since
their weights are far above the forget threshold.

On the other hand, the mean edge weight in BA simply diminishes its decreasing
rate, until it starts to fluctuate around a stabilisation value. This value is
low, highlighting the fact that, on average, the semantic networks built using
BA are ``weak". In fact, due to their low weight value, edges can be more
subject to be forgotten.

\subsubsection{Scenario 2}

\begin{figure}[htbp] \centering
  \subfloat[\label{fig:comp150KD2}]
  {\includegraphics[width=0.495\columnwidth]{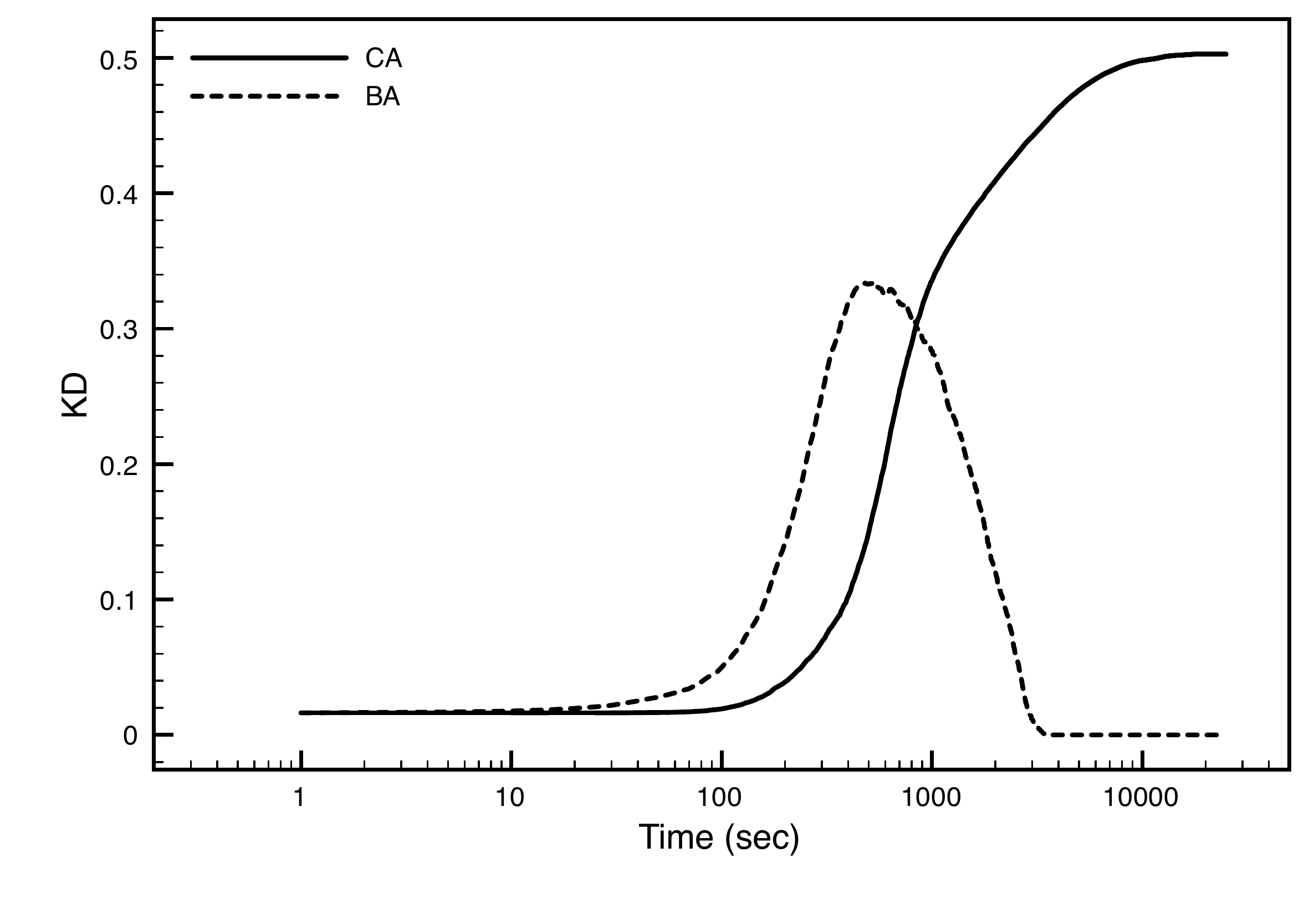}}
  \subfloat[\label{fig:comp150DE2}]
  {\includegraphics[width=0.495\columnwidth]{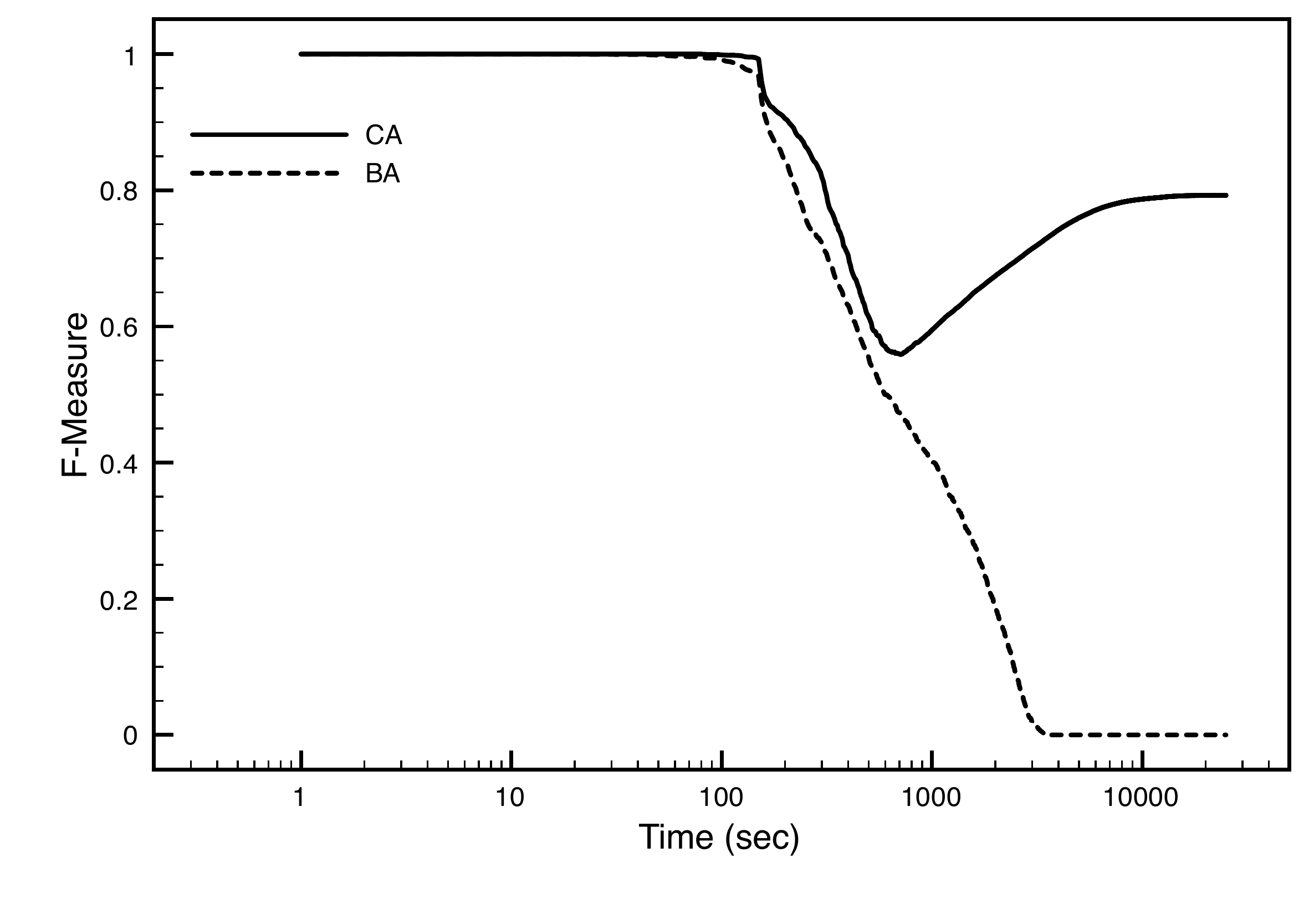}}
  \caption{KD(a) and F-measure(b) comparison with $f_{min}=150s$ on dataset $D2$
  (Scenario 2).
  \label{fig:comp150d2}}
\end{figure}

\begin{figure}[htbp] \centering
  \subfloat[\label{fig:comp300KD2}]
  {\includegraphics[width=0.485\columnwidth]{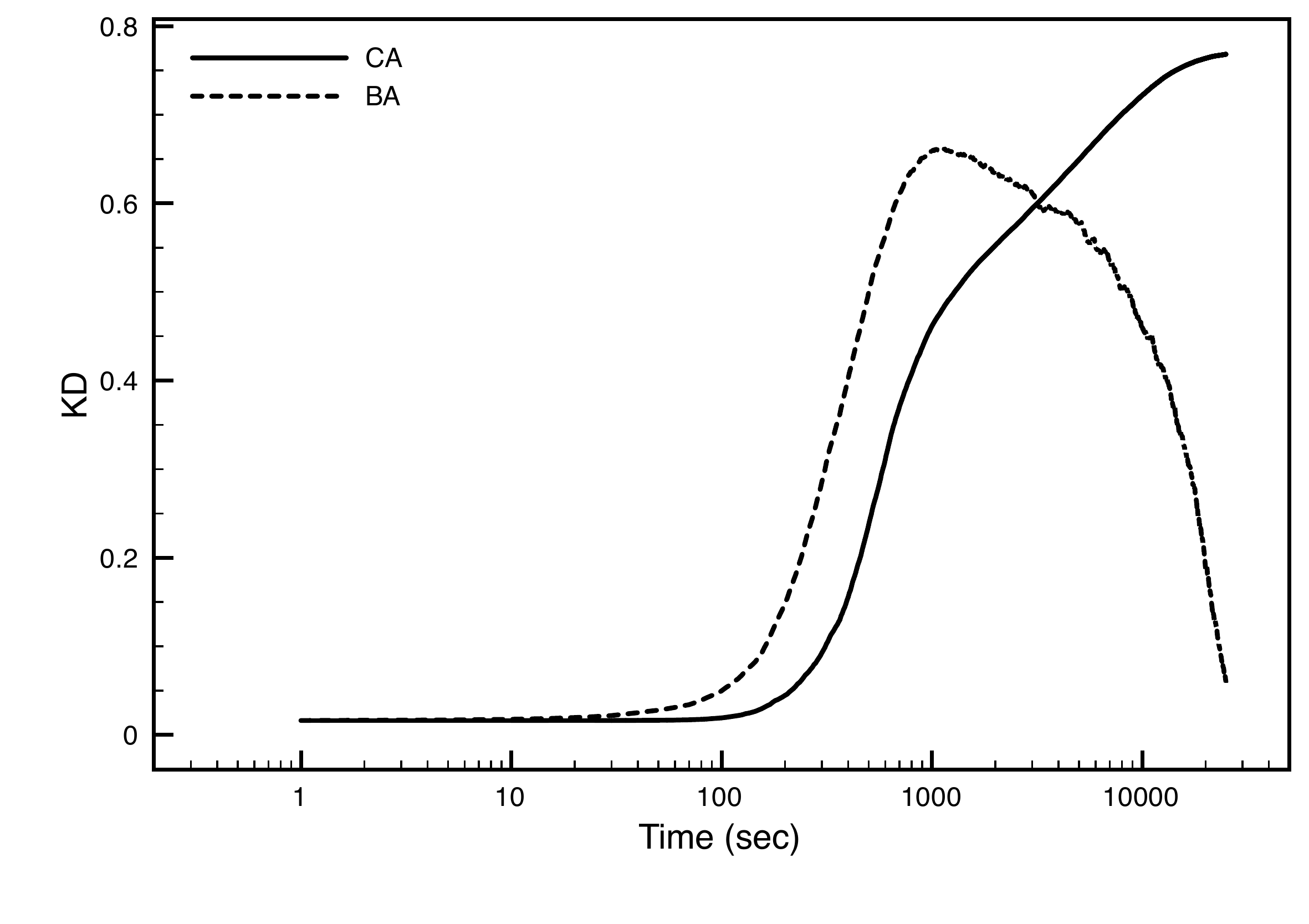}}
  \subfloat[\label{fig:comp300DE2}]
  {\includegraphics[width=0.485\columnwidth]{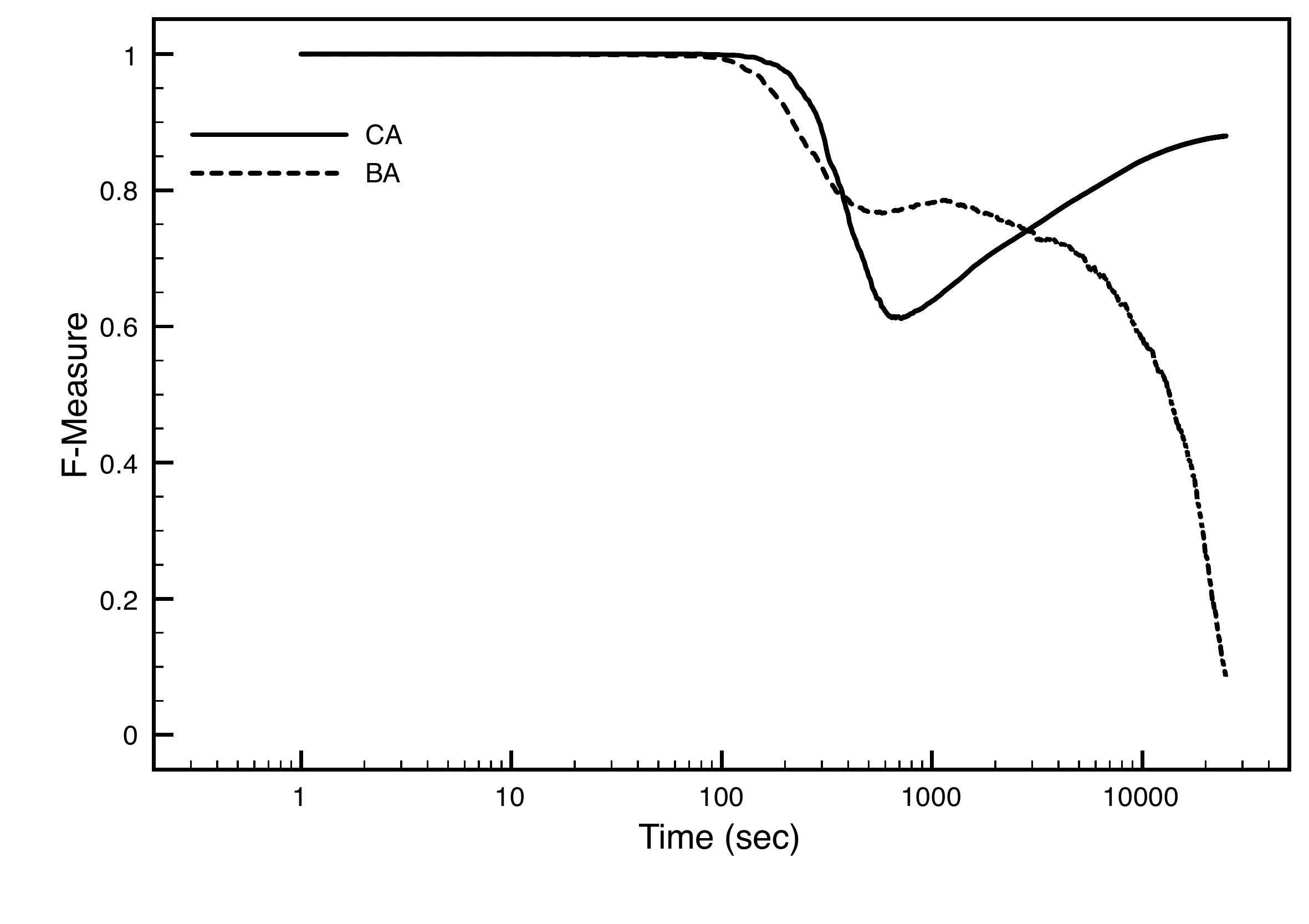}}
  \caption{KD(a) and F-measure(b) comparison with $f_{min}=300s$ on dataset $D2$
  (Scenario 2).
  \label{fig:comp300d2}}
\end{figure}

In order to further analyse the behaviour of the two
approaches, we now show results obtained using dataset {\em D2}.
Figs.~\ref{fig:comp150d2}--\ref{fig:comp300d2} refer to the single-community
scenario defined in {\em Scenario 2}.

In both these cases, it is possible to note the same effects observed with
dataset {\em D1}.  The F-measure of the CA approach initially decreases. After
this phase, the semantic knowledge owned by each device starts to attract more
and more related data items, leading to an increase of the F-measure metric. For
BA, the starting delay on  data dissemination is successively worsened by the
lost of semantic information, leading to a continuous decrease of F-measure
values.

With respect to the KD metric, BA shows an initial faster growth of the semantic
networks, but it is not able to preserve the retrieved information over time.
With $f_{min}=150s$ (Fig.~\ref{fig:comp150d2}), BA reaches KD = 0, and
also with $f_{min}=300s$ (Fig.~\ref{fig:comp300d2}) it gets
close to  this value. In all these cases, CA proceeds initially
slower, allowing its semantic networks to strengthen their structure, finally
achieving high values of KD.

The difference in the knowledge acquisition process are also reflected in
quality of the retrieved data. This measure is also affected by the

approach used by BA to select the data to be exchanged upon contact. Note that,
with $f_{min}=150s$, the BA approach is never able to achieve better values of
the F-measure than CA. With $f_{min}=300s$, BA only shortly has a F-measure
better than CA, eventually declining the values of this metric, in parallel with
the loss of semantic knowledge.

\subsubsection{Scenario 3}

\begin{figure}[htbp]
  \centering
  \subfloat[\label{fig:comp150KD}]
  {\includegraphics[width=0.495\columnwidth]{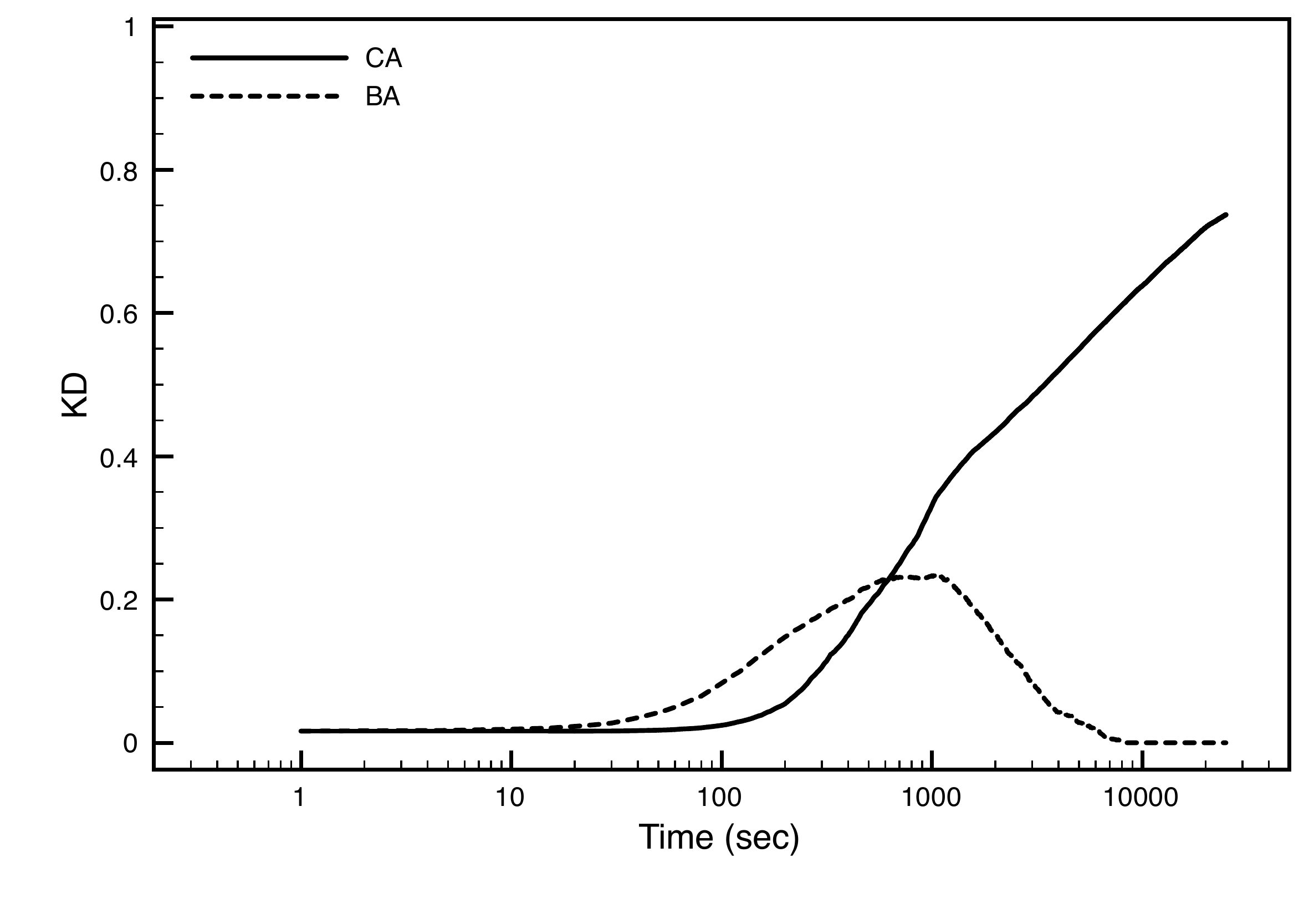}}
  \subfloat[\label{fig:comp150DE}]
  {\includegraphics[width=0.495\columnwidth]{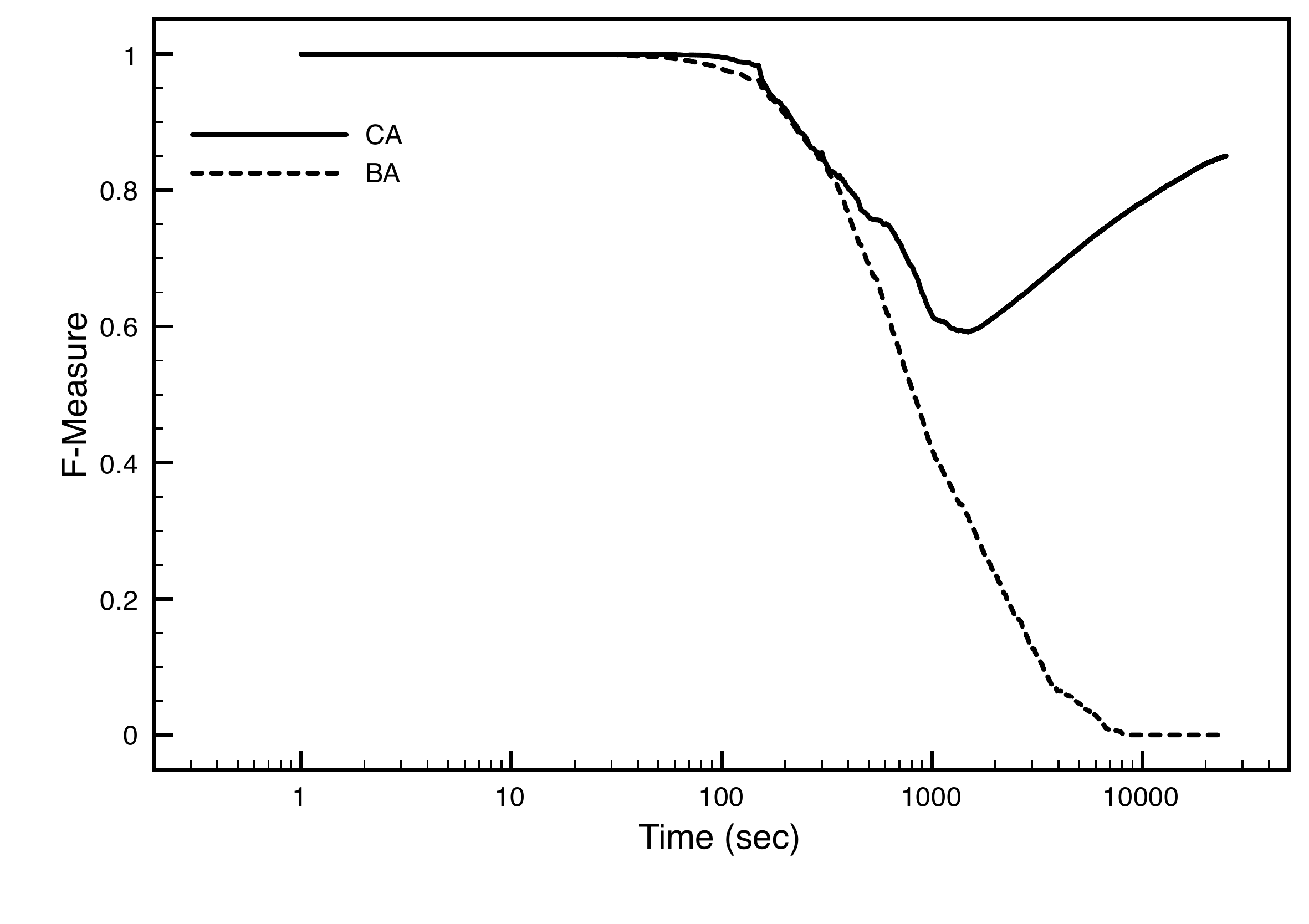}}
  \caption{KD(a) and F-measure(b) comparison with $f_{min}=150s$ on dataset $D1$
  (Scenario 3).
  \label{fig:comp150d2-3comm}}
\end{figure}

In the last scenario, we tested the two competing
approaches in a more complex situation. Nodes are divided into three separated
communities, and are equipped with data coming from the {\em D1} dataset.
Non-travelling nodes have to rely on travellers to collect both knowledge and
data spread in the other communities. Fig.~\ref{fig:comp150d2-3comm} shows the
variation of KD and F-measure with $f_{min}=150s$. In this case, the usual
initial advantage of BA over CA in the KD metric lasts for a very short time. In
the F-measure metric, BA starts to decline, without ever stopping . 

With $f_{min}=300s$, BA shows an increased ability to preserve the content of
the semantic networks of the system. However, BA eventually enters a declining
phase, although this process is slower than in all
the previous cases. In order to better
investigate the behaviour of BA (and CA) in this case, we allowed the simulation
to run longer, i.e. for 125,000 sec. It is possible to observe that the declining phase of
the BA curve leads the CA solution to outperform BA, as in all the other cases.
This fact can be observed for the F-measure, too. Initially, BA has a better
performance, but it finally starts decreasing. At the
same time, the F-measure curve of CA increases, and CA ends the simulation with a better performance than BA.

%This is slower that those of all the previous cases, but it is anyway present.
%On the other hand, CA ends the simulation in a clear increasing phase w.r.t.
%the size of its semantic networks. It is reasonable to suppose that, with
%enough time available CA will outperform BA also in this case. As for the
%F-measure, at the end of the simulation, BA and CA have almost the same value.
%However, as in the KD metric case, BA shows a decreasing trend of its
%F-measure, while CA keeps increasing the value of this metric. Therefore, also
%for this case, it is reasonable to suppose that CA will finally outperform BA. 

\begin{figure}[htbp]
  \centering
  \subfloat[\label{fig:comp300KD}]
  {\includegraphics[width=0.495\columnwidth]{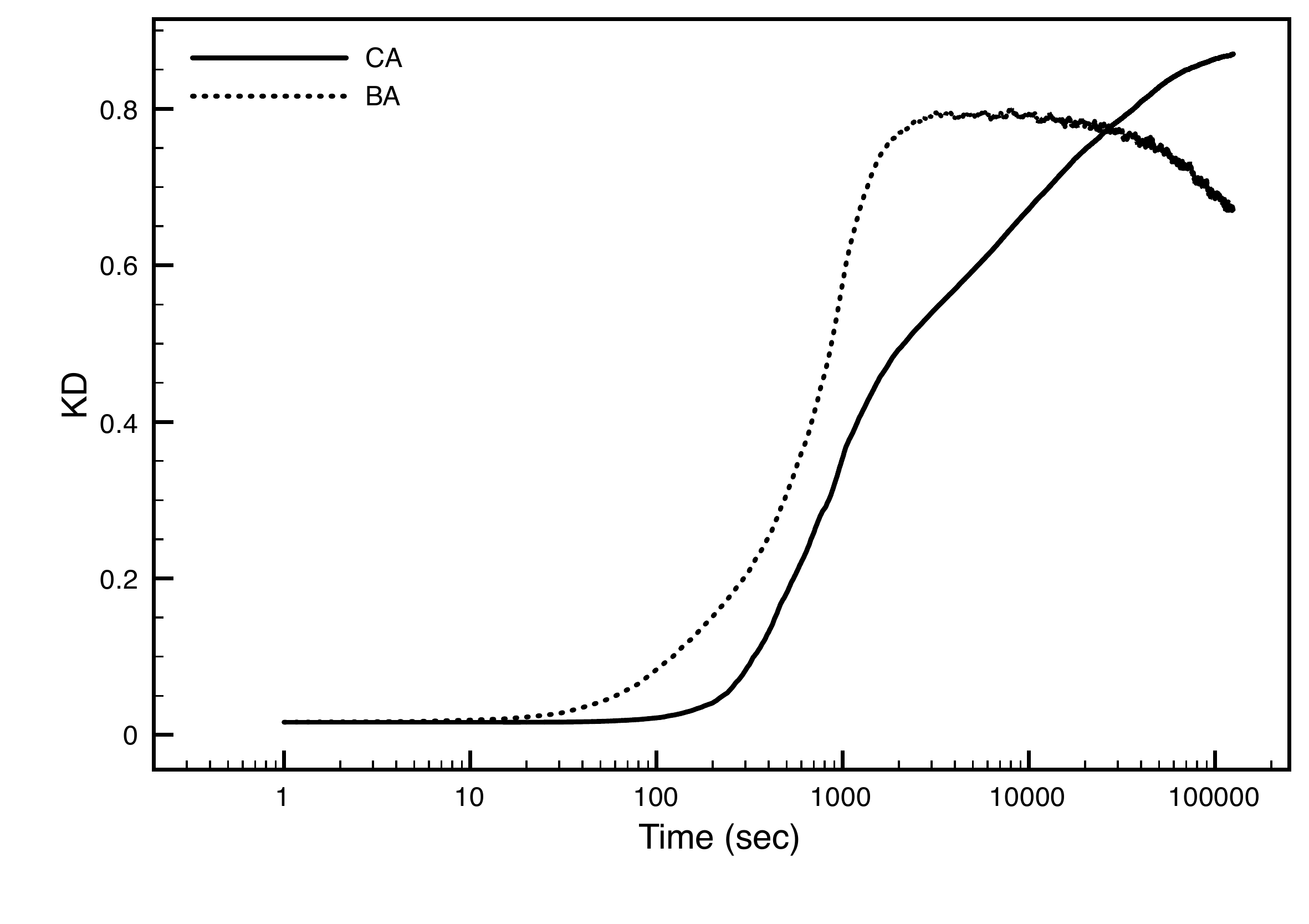}}
  \subfloat[\label{fig:comp300DE}]
  {\includegraphics[width=0.495\columnwidth]{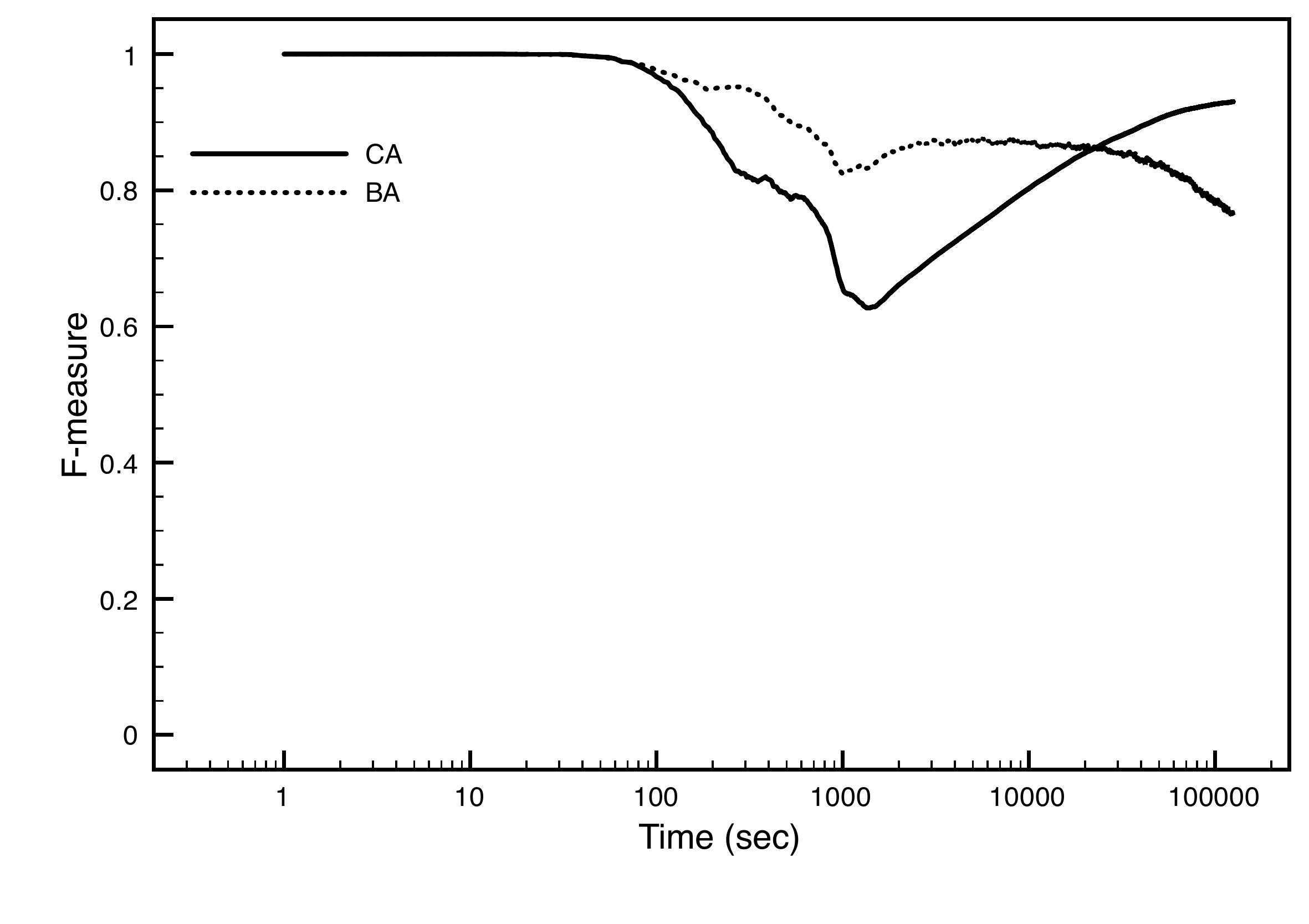}}
  \caption{KD(a) and F-measure(b) comparison with $f_{min}=300s$ on dataset $D1$
  (Scenario 3).
  \label{fig:comp300d2-3comm}}
\end{figure}

\subsubsection{Coverage}
\label{sssec:coverage}
Tab.~\ref{tab:convCov} presents the performance under all the scenarios of both CA and BA with respect to the
Coverage metric. The table reports the values and the time instants at which Coverage stabilises. Specifically, convergence is defined as the point in time, during a simulation, when the coverage value does not change any more until the end of the simulation. Note that, as data items are not dropped by nodes and no new data items are generated during simulations, coverage can only increase. It stops increasing, and therefore remains constant, when the dissemination process becomes stable and nodes do not exchange any more data items during contacts. Consequently, according to this very strict definition of coverage,  convergence times may appear quite high. However we decided to use this definition because we are interested in studying the performance of our system when the dissemination process enters in steady state.

Generally, in almost all the cases, convergence is reached by CA faster, and
with higher Coverage values, than BA. In particular, it is possible to note that, for
$f_{min} = 150s$, BA ends up with a Coverage of 0. This is due to the deletion
of all the semantic concepts from the nodes' semantic networks, as already
observed in all the previous experiments. In only one case, for $f_{min}=300s$
under {\em scenario 2}, BA is able to reach convergence faster than
CA.
~In this case, the  $f_{min}$ value gives time to BA to rapidly accumulate a high number of data items. Given the fact that data items are never discarded once acquired, the definition of the Coverage measure (see  Formula~\ref{eq:coverage}) implies that the Coverage value could remain high even when the semantic knowledge starts to decrease, due to forgetting. However, all the other metrics (see Fig.~\ref{fig:comp300d2}) highlight the fact the performance of the system is degrading.
 From these observations, we can reasonably suppose that
also in this case the Coverage metric will start to decrease, possibly reaching
0.

\begin{table}[ht]
  \centering
  \caption{Convergence values and times for the Coverage metric in all the
    considered scenarios
    \label{tab:convCov}}{
      \renewcommand{\arraystretch}{1.2}
      \begin{tabular}{cc|c|c|c|c|c|c|}
	\cline{3-8} & & \multicolumn{2}{|c|}{\emph{Scen. 1}} &
	\multicolumn{2}{|c|}{\emph{Scen. 2}} & \multicolumn{2}{|c|}{\emph{Scen. 3}}  \\
	\cline{3-4} \cline{2-8} &  \multicolumn{1}{|c|}{$f_{min}$}  &
	{\em CVG} & $t$ &  {\em CVG} & $t$ &  {\em CVG} & $t$ \\
	\hline
	\multicolumn{1}{|c|}{\multirow{2}{*}{CA} }& 150 & 0.987 & 2105 & 0.977 & 22895 & 0.993 & 2605\\
	\cline{2-8} \multicolumn{1}{|c|}{} & 300 & 0.989 & 2160 & 0.997 & 24970 & 0.999 & 2605 \\
	\hline\hline
	\multicolumn{1}{|c|}{\multirow{2}{*}{BA} }& 150  & 0.0 & 3635 & 0.0 & 24360 & 0.0 & 8295\\
	\cline{2-8} \multicolumn{1}{|c|}{} & 300 & 0.599 & 24925 & 0.999 & 8040 & 0.999 & 23255\\
	\hline
      \end{tabular}}
\end{table}

As a general remark on these results, we can note that CA achieves good Coverage
values even with a low $f_{min}$. This implies that, even in face of a limited
resource consumption (low $f_{min}$), CA is able to let relevant data items flow
toward interested nodes in the system.

\subsection{Analysis on node's semantic networks}
\label{ssec:graphAnalysis}

Next, we analyse in more detail the nodes' semantic networks for both the approaches. For the sake
of simplicity,  the results presented hereafter will refer to a tagged node that
we can consider, without loss of generality, as representative of all the other
nodes in the system. Specifically, for Scenarios like 1 or 2 this is any
  of the nodes in the only existing community. For Scenario 3, we consider one
  of the non-traveller nodes.

\begin{figure}[ht]
  \centering
  \includegraphics[width=.6\textwidth]{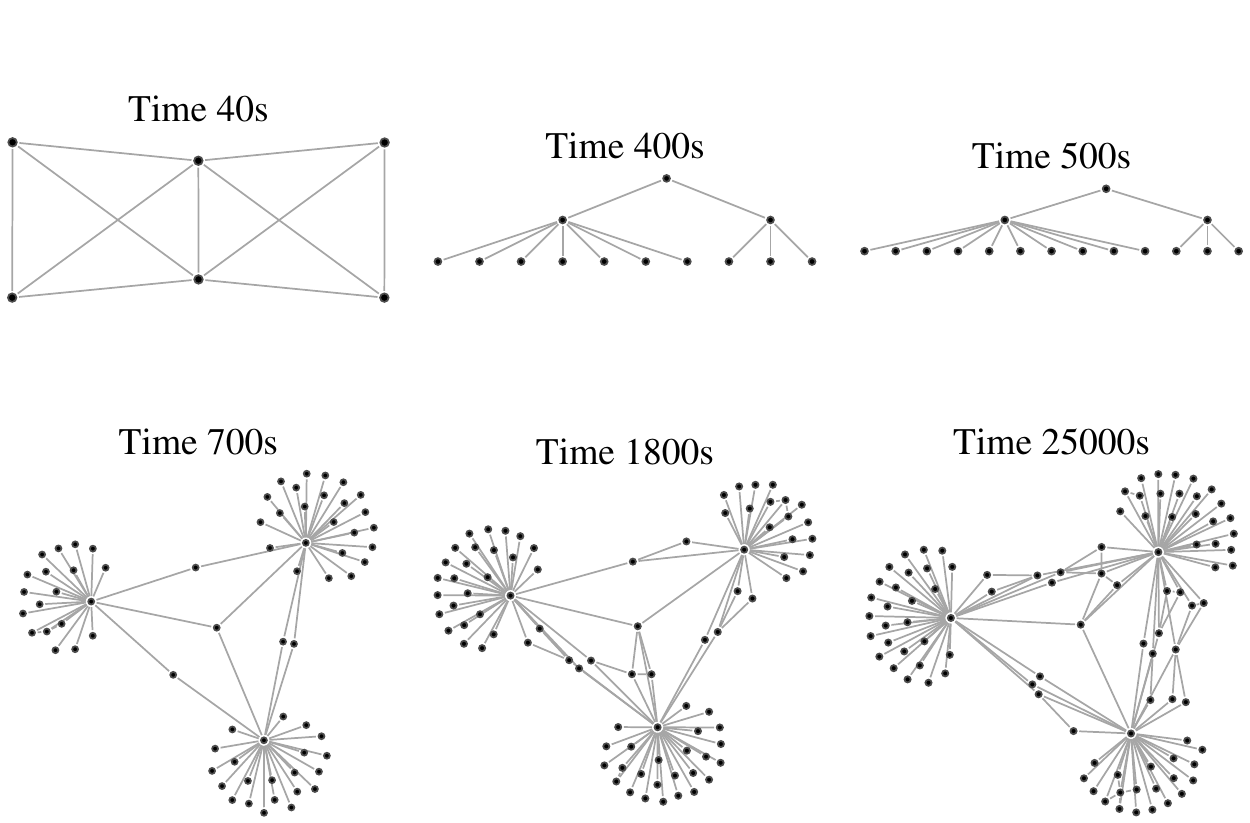}
  \caption{Evolution of a tagged node's semantic network evolution running CA in
  Scenario 1.}
  \label{fig:CASnEvol}
\end{figure}

\begin{figure}[ht]
  \centering
  \includegraphics[width=.6\textwidth]{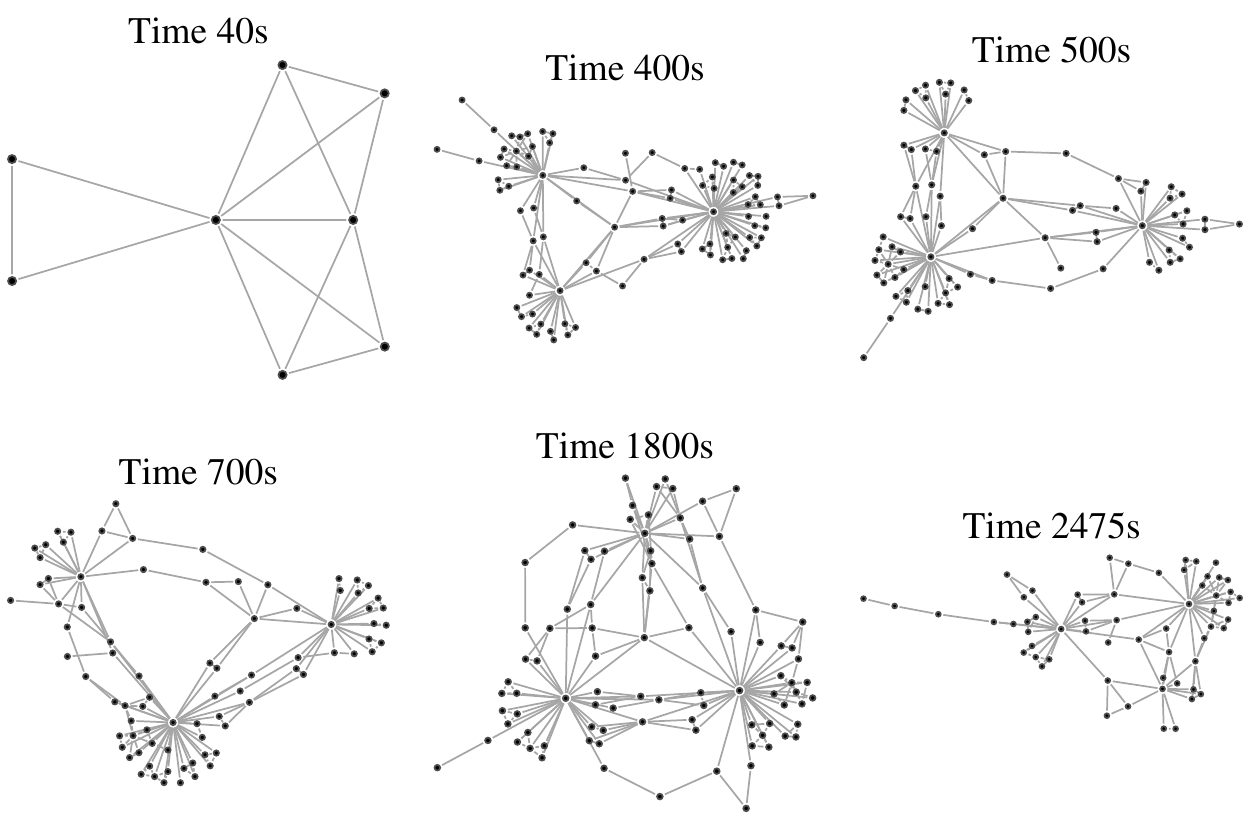}
  \caption{Evolution of a tagged node's semantic network evolution running BA in
  Scenario 1.}
  \label{fig:BASnEvol}
\end{figure}

\begin{figure}[ht]
  \centering
  \includegraphics[width=.735\textwidth]{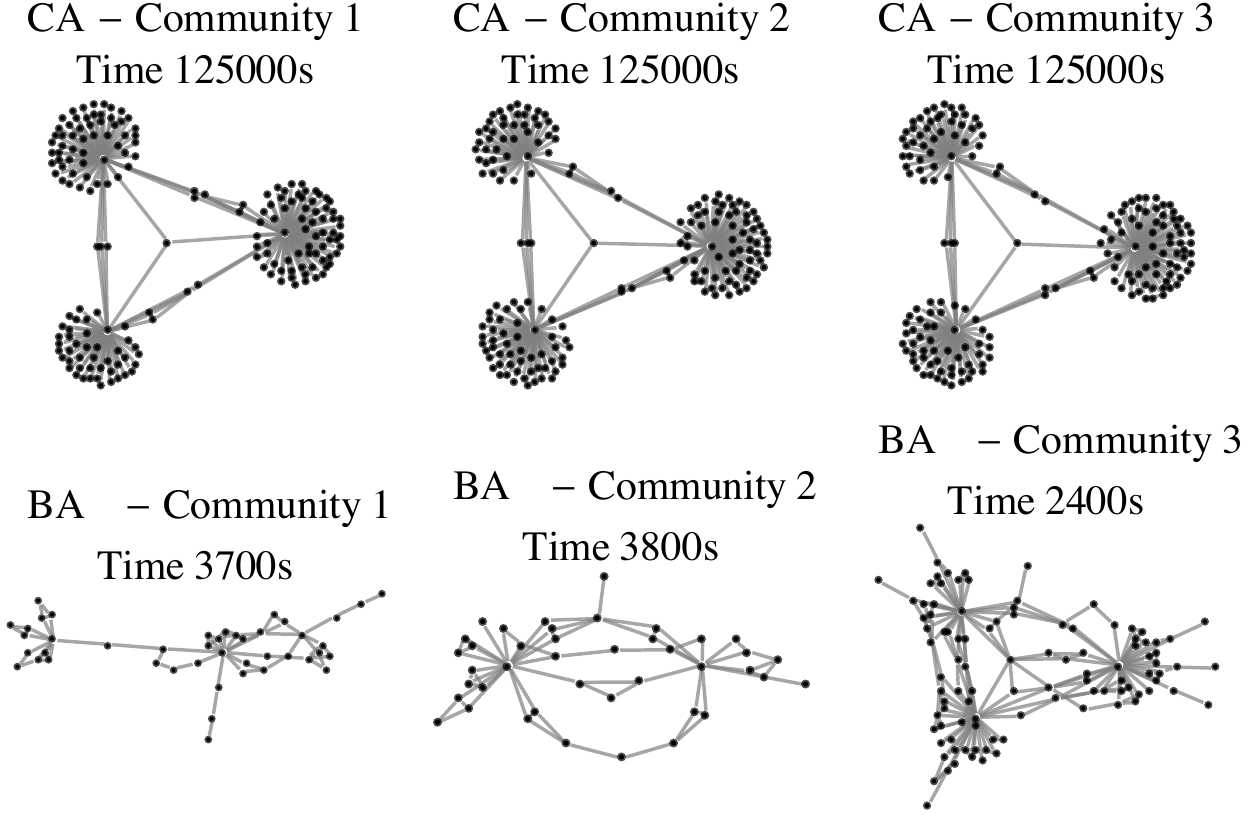}
  \caption{Final configuration of three tagged nodes' SNs running CA (upper
  half) and BA (lower half) in Scenario 3.}
  \label{fig:3CommCompEvol}
\end{figure}

We tracked the state of the tagged node's SN at
successive time instants.  Figs. \ref{fig:CASnEvol} and \ref{fig:BASnEvol} refer
to the SN evolution in CA and BA, respectively. We
can again observe the different growth rate of the SN triggered by the two
approaches. With CA, the tagged node's SN grows more slowly than with BA,
however this leads to a more robust SN evolution than with BA.
If we compare the  state of the SNs at $700s$ we
notice that the structure of the
SN built by CA is more similar to $G1$ than the one built according to  BA.
Moreover,  despite the faster initial knowledge acquisition of BA, the
randomness of the mechanism leads to a more chaotic SN structure, characterised
by long and weak paths between vertexes.  This behaviour makes the resulting SN
very unstable and more susceptible to the forgetting process. For
example, when using BA, after
$2475s$ all the semantic information in the  SN has been lost. Conversely,
thanks to the cognitive process, CA tends to attach the new information around
an increasingly stable core. The very same behaviour can be also observed in the
multi-community environment of {\em Scenario 3}, as shown in
Fig.~\ref{fig:3CommCompEvol}. The upper part of the figure reports the final
configuration of the SNs of three tagged nodes (one per community)
using CA. The
other half of the  figure presents the configurations of  the SNs of the 3
tagged nodes when the BA scheme is used. The latter configurations are taken
just before the SNs of each node become empty, due to the forgetting process.
%LORENZO: TODO check se anche in scenario 3 abbiamo lo stesso comportamento. 

Also in this case, the cognitive mechanisms
of the CA
scheme lead the SN structure to
be very similar to $G1$. It is worth noting that no node in the system knows the
real structure of $G1$, but nevertheless, they are able to replicate a structure
very similar to the global one, that could be in principle obtained only
with global knowledge. Another key feature of CA is that at the end
of the simulation, all nodes' SN are very similar. This can be seen as some kind of a
spontaneous consensus. That is, in a closed environment where no new information
is generated and when information can eventually circulate across all
nodes, knowledge tend to become homogeneous among nodes, which align
towards a common representation of available semantic information.

This kind of similarity is reflected by structural properties of the nodes' SNs.
We averaged over 10 simulated experiments  the tagged node's final semantic
network size for both CA and BA and compared it with the
complete sizes of the graphs ($G1$ and $G2$).  
%Notice that, $G2$ is much greater than $G1$ in size then, the nodes' SNs in
%Scenario 2, although smaller in percentage than the ones in Scenarios 1,3 have
%the same order of magnitude.  LORENZO TODO rifrasare la frase commentata e
%renderla meno oscura
Table~\ref{tab:snpropperc} shows the number of vertices and edges in the
final SN of the tagged node as a fraction of those in the global graphs.
Diameter values are absolute. In this case, diameters of $G1$ and $G2$ are
$4$ and $8$, respectively.
As reported in the Table, though the SN
formed with CA is
smaller than the corresponding global graph,
it approximates very well the
 diameters of the global graphs.   On the
other hand, BA is not as accurate as CA. Most of the time BA is not able to
preserve the information in the SN until the end of simulation. Only when the
forgetting time is sufficiently high the information lasts until the end
(Scenario 2,3 with  $f_{min}=300$) and even in that case the graph properties
are not always respected (e.g., see the much different diameter values).

\begin{table}[bt]
  \begin{center}
    \caption{Node's final semantic network properties for CA and BA. Mean values
      and standard deviations computed on $10$ runs are reported.
      Missing values (``$-$'') means that nodes' semantic network are empty, as a
      consequence of the forgetting process.
      \label{tab:snpropperc}}{
	\begin{tabular}{|c|c||c|c|c||c|c|c|}
	  \hline
	  &  & \multicolumn{3}{c||}{\textit{CA}} & \multicolumn{3}{c|}{\textit{BA}}
	  \\ \cline{3-8} \textbf{Scenario} &$f_{min}$& \textbf{\% Edges} &
	  \textbf{\% Vertex} & \textbf{Diameter} & \textbf{\% Edge} & \textbf{\%
	  Vertex} & \textbf{Diameter}  \\
	  \hline
	  \multirow{2}{*}{1} & $150$ & $38.5\pm3.9$ & $71.3\pm6$ & $4\pm0.0$ & $-$ & $-$ & $-$ \\
	  & $300$ & $39.7\pm5.9$ & $59\pm6.7$ & $4.1\pm0.3$  & $-$ & $-$ & $-$ \\
	  \hline
	  \multirow{2}{*}{2} & $150$ & $5.86\pm2.1$ & $18.3\pm6.2$ & $6.3\pm0.5$
	  & $-$ & $-$ & $-$\\
	  & $300$ & $9.6\pm2.3$ & $22.7\pm4.9$ & $6.2\pm0.4$  & $7.8\pm2.3$ & $26.2\pm4.5$ & $15\pm2.6$ \\
	  \hline
	  \multirow{2}{*}{3} & $150$ & $41\pm4.6$ & $64.8\pm5.4$ & $4\pm0.0$ &
	  $-$ & $-$ & $-$ \\
	  & $300$ & $56.8\pm5.7$ & $67.7\pm5.3$ & $4\pm0.0$
	  & $91.6\pm3.5$ & $88.3\pm4.4$ &  $4.6\pm0.7$ \\
	  \hline
	\end{tabular}
      }
    \end{center}
\end{table}

In order to deeply investigate the similarity between the final SN
generated by CA and BA, and
the original graphs G1 and G2, we analyse the vertex degree
distribution of the SNs. Namely, we compared the empirical vertex degree
distribution of the node's final SN obtained with CA and BA with the empirical
distribution of G1 and G2. 
%We precise that the empirical distribution for G1 and G2 are computed
%considering the $99.8$-th percentile of the each dataset. 
We used the Cramer-von-Mises goodness of fit test with a
significance level of $95\%$ to compare the distributions.   Results are reported in Table
\ref{tab:cvm}.

\begin{table}[ht]
  \begin{center} \caption{Cramer von Mises
      goodness of fit test computed on the degree distribution of a tagged node
      semantic networks obtained with CA and BA. Results have been computed on
      $10$ runs. R and NR mean ``Reject'' and ``Do not reject'',
      respectively.
      \label{tab:cvm}}
      {
	\begin{tabular}{|c|c||c|c|c||c|c|c|}
	  \hline
	  &  &
	  \multicolumn{3}{c||}{\textit{CA}} & \multicolumn{3}{c|}{\textit{BA}}
	  \\ \cline{3-8} \textbf{Scenario} &$f_{min}$& \textbf{Stat.} &
	  \textbf{P-Value} & \textbf{Concl.} & \textbf{Stat.} & \textbf{P-Value}
	  & \textbf{Concl.} \\
	  \hline
	  \multirow{2}{*}{1} & $150$ & $0.6225 $ & $0.01970$ & R & $-$ & $-$ & $-$ \\
	  & $300$ & $0.3925$ & $0.0756$ & \textbf{NR}  & $-$ & $-$ & $-$ \\
	  \hline
	  \multirow{2}{*}{2} & $150$ & $0.0679606$ & $0.764137$ & \textbf{NR} & $-$ & $-$ & $-$ \\
	  & $300$ & $0.157223$ & $0.368298$ & \textbf{NR}  & $5.004$ & $2.99261*10^{-12}$
	  & R \\
	  \hline
	  \multirow{2}{*}{3} & $150$ & $0.385$ & $0.0791754$ &
	  \textbf{NR} & $-$ & $-$ & $-$ \\
	  & $300$ & $0.2325$ & $0.212442$ &
	  \textbf{NR}  & $0.1825$ & $0.304102$ & \textbf{NR} \\
	  \hline
	\end{tabular}
      }
    \end{center}
\end{table}

\begin{figure}[ht]
  \centering
  \subfloat[CA]{ \includegraphics[width=.435\textwidth]{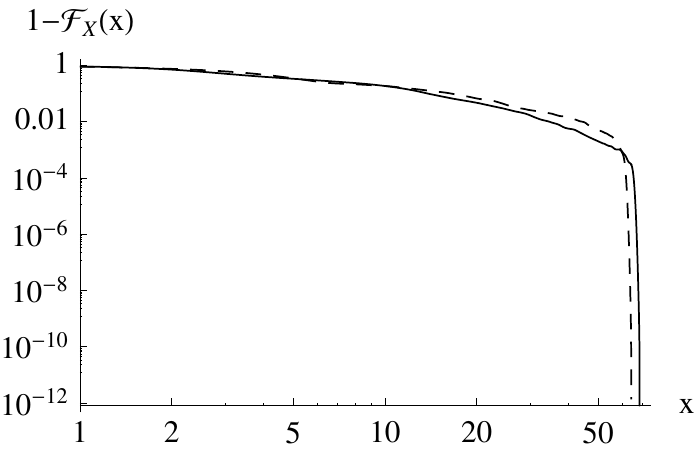}
  \label{fig:ccdfCA} }
  \subfloat[BA]{\includegraphics[width=.435\textwidth]{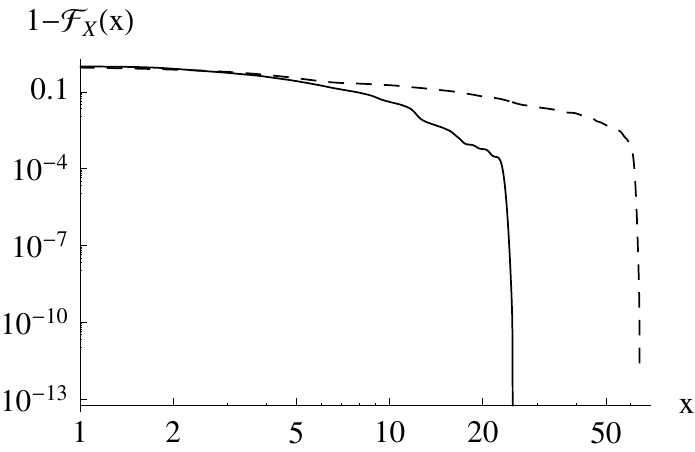}
    \label{fig:ccdfBA} }
  \caption{CCDFs of the degree distribution for CA and BA
      (solid curves) compared with the degree distribution of $G2$ (dashed
      curve).\label{fig:dist_comp}}
\end{figure}

As we can see, in almost all the experiments the empirical degree distribution 
% (noted as $\mathcal{\widetilde D}$)  
of node's SN using CA is the same as that of the global
graphs.  Differently, BA only in one case was able
to reproduce a structure  similar to  the original
graph. As an example, figures
\ref{fig:ccdfCA}-\ref{fig:ccdfBA} show the comparison between the CCDF of the
degree distribution of the final SN with the one of G2 (Scenario
2, $f_{min}=300$), both for CA and BA.

We point out that for CA the degree distribution of nodes' SN is equivalent to
the global graph, up to a constant shift.  

%\todolnLV{Effettivamente pensandoci meglio il concetto di shift non ha molte senso stressarlo.}

%
% to the  empirical degree distribution of the original graph.
% $$\mathcal{\widetilde D}_{SN} \equiv \mathcal{ \widetilde D}_G - c$$ where
% $c=\min\mathcal{\widetilde{D}_G}-\min \mathcal{\widetilde D_{SN}}$.
%\begin{table}[ht] \begin{center} \tbl{Node's final semantic network properties
%for CA and BA. Mean values with confidence interval at $95\%$ computed on $10$
%run are reported. \label{tab:snprop}}{ \begin{tabular}{|c|c||c|c|c||c|c|c|}
%\hline &  & \multicolumn{3}{c||}{\textit{CA}} &
%\multicolumn{3}{c|}{\textit{BA}} \\ \cline{3-8} \textbf{Scenario} &$f_{min}$&
%\textbf{Edges \#} & \textbf{Vertex \#} & \textbf{Diameter} & \textbf{Edge \#} &
%\textbf{Vertex \#} & \textbf{Diameter}  \\ \hline \multirow{2}{*}{1} & $150$ &
%$175[13]$ & $157[10]$ & $4[0]$ & $-$ & $-$ & $-$ \\ & $300$ & $180[18]$ &
%$130[10]$ & $4.1[0.22]$  & $-$ & $-$ & $-$ \\ \hline \multirow{2}{*}{2} & $150$
%& $186[18]$ & $143[8]$ & $4[0]$ & $-$ & $-$ & $-$ \\ & $300$ & $258[18]$ &
%$150[8]$ & $4[0]$  & $416[11]$ & $195[7$ & $4.6[0.5]$ \\ \hline
%\multirow{2}{*}{3} & $150$ & $518[132]$ & $239[57]$ & $6.3[0.34]$ & $-$ & $-$ &
%$-$ \\ & $300$ & $855[150]$ & $295[45]$ & $6.2[0.3]$  & $694[143]$ & $241[42]$
%& $15[2]$ \\ \hline \end{tabular} } \end{center}��� \end{table}
%

\subsection{Sensitivity analysis}
\label{ssec:sens}

Having compared in detail CA and BA, in this section we analyse the impact of the CA parameters on its behaviour. The following results are obtained under the conditions of {\em Scenario 1}.

\begin{figure}[ht!] \centering
  \includegraphics[width=20em]{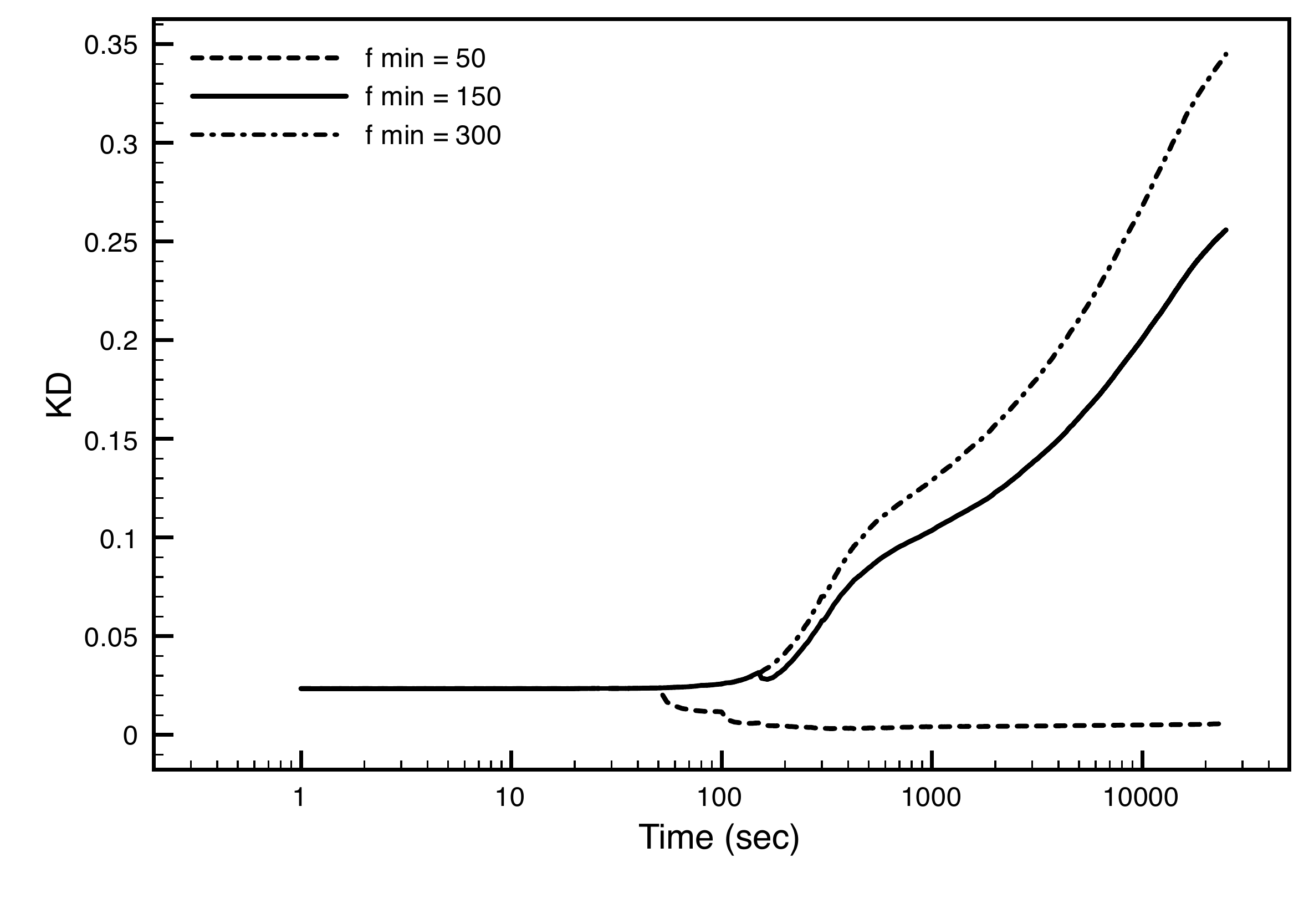}
  \caption{Impact on KD of the {\em remember} threshold} \label{fig:sensForget}
\end{figure}

In Fig.~\ref{fig:sensForget}
we  show how the KD metric varies by using three
different settings of the \emph{forget} threshold value. As expected, higher
values of $f_{min}$ (i.e. edges are ``remembered" for longer times) allow the
system to achieve higher values of KD. In fact, less popular edges can be kept
in each node's semantic network for longer periods, and, thus, they can have
more chances to be exchanged during contacts with other peers. However, the
lowest value of the \emph{forget} threshold highlights the fact that, with a
very rapid forgetting process, a phase transition could happen. In fact, in this
case, rather than observing an increase in the KD  metric, nodes in the system
experience a loss of semantic information in their semantic networks.
In other words, for values of the \emph{forget} threshold below the one
marking the phase transition,  also CA looses information faster than it can
consolidate it. Beyond this value, information has enough time to consolidate in
the semantic networks, and does not disappear from the network. 
 
\begin{figure}[ht!]
  \centering
  \includegraphics[width=20em]{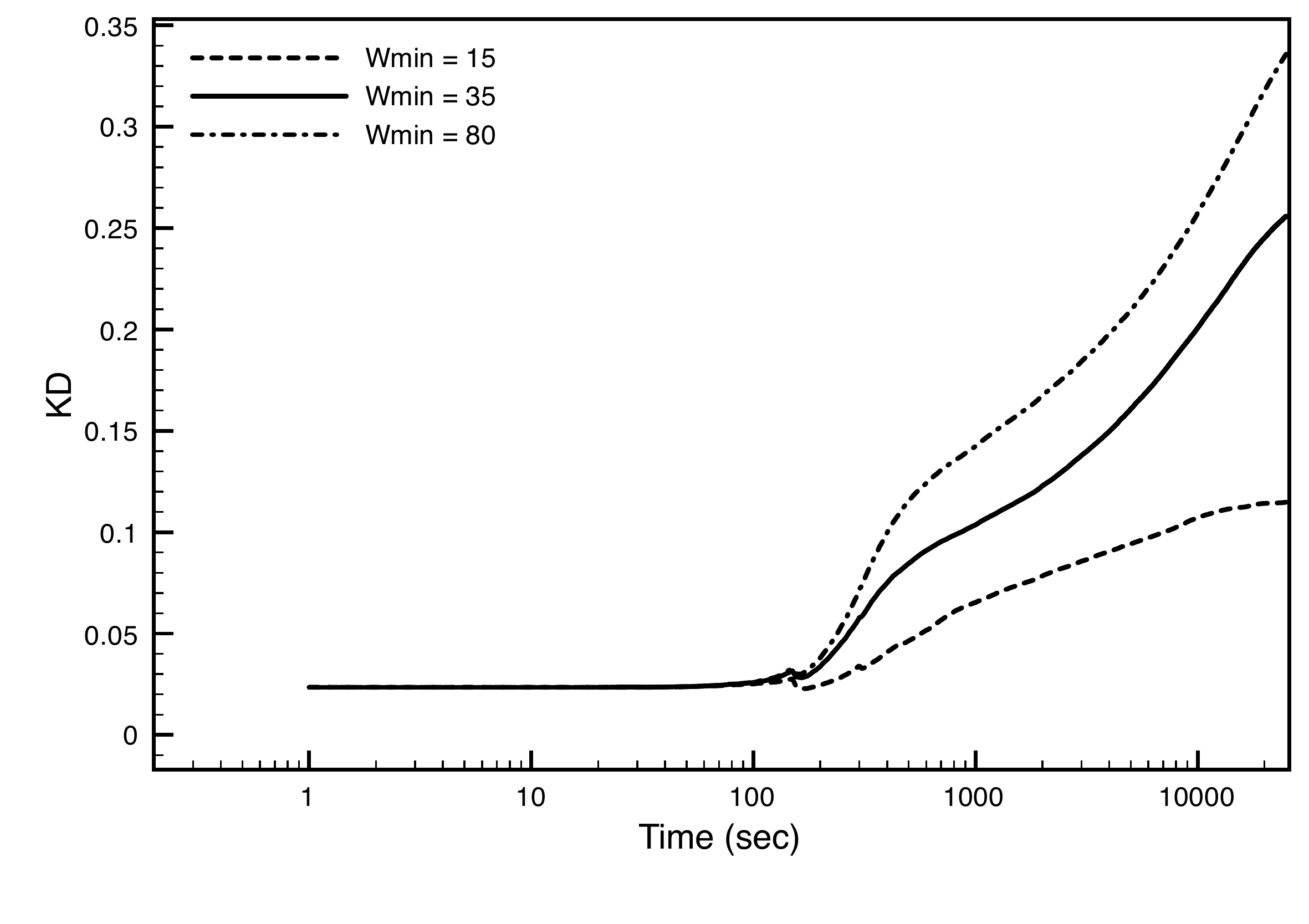}
  \caption{Impact on KD of the {\em weight} threshold}
  \label{fig:sensWeight}
\end{figure}

Fig.~\ref{fig:sensWeight} presents the performance of the system with different
values of the $W_{min}$ threshold (remember that the higher $W_{min}$ the
more edges can be included in the semantic network exploration process between two nodes
during a contact). Also in this case, higher values of the
threshold allow the system to achieve higher values of KD. However, there
  is a sort of marginal utility gain in increasing the value of $W_{min}$, which
means that increasing the size of the semantic network explored during a
contact is not worth after a certain point. 

%A very low value of $W_{min}$ limit the possibility of a node to continue the
%exploration of its semantic network far from key vertices and inhibiting to
%follow ``aged" (i.e. not activated for long time)  edges.  Therefore, being
%very restrictive on the possibility to explore the donor network (i.e. setting
%very low values of $W_{min}$) has a bigger impact on the final performance of
%the system, than the one existing by different, yet high, values of the
%threshold.  
%
%Io leggerei questi risultati in modo un po' diverso. C'e' un fenomeno di
%marginal utility: quando aumento W da 15 a 35 ho un aumento maggiore che da 35
%ad 80. Significa che il vantaggio marginale di esplorare la donor network piu'
%estensivamente decresce via via, quindi non e' cosi' importante esplorare la
%rete molto estensivamente lontano dai common vertices.

\begin{figure}[ht!] \centering
  \includegraphics[width=19.75em]{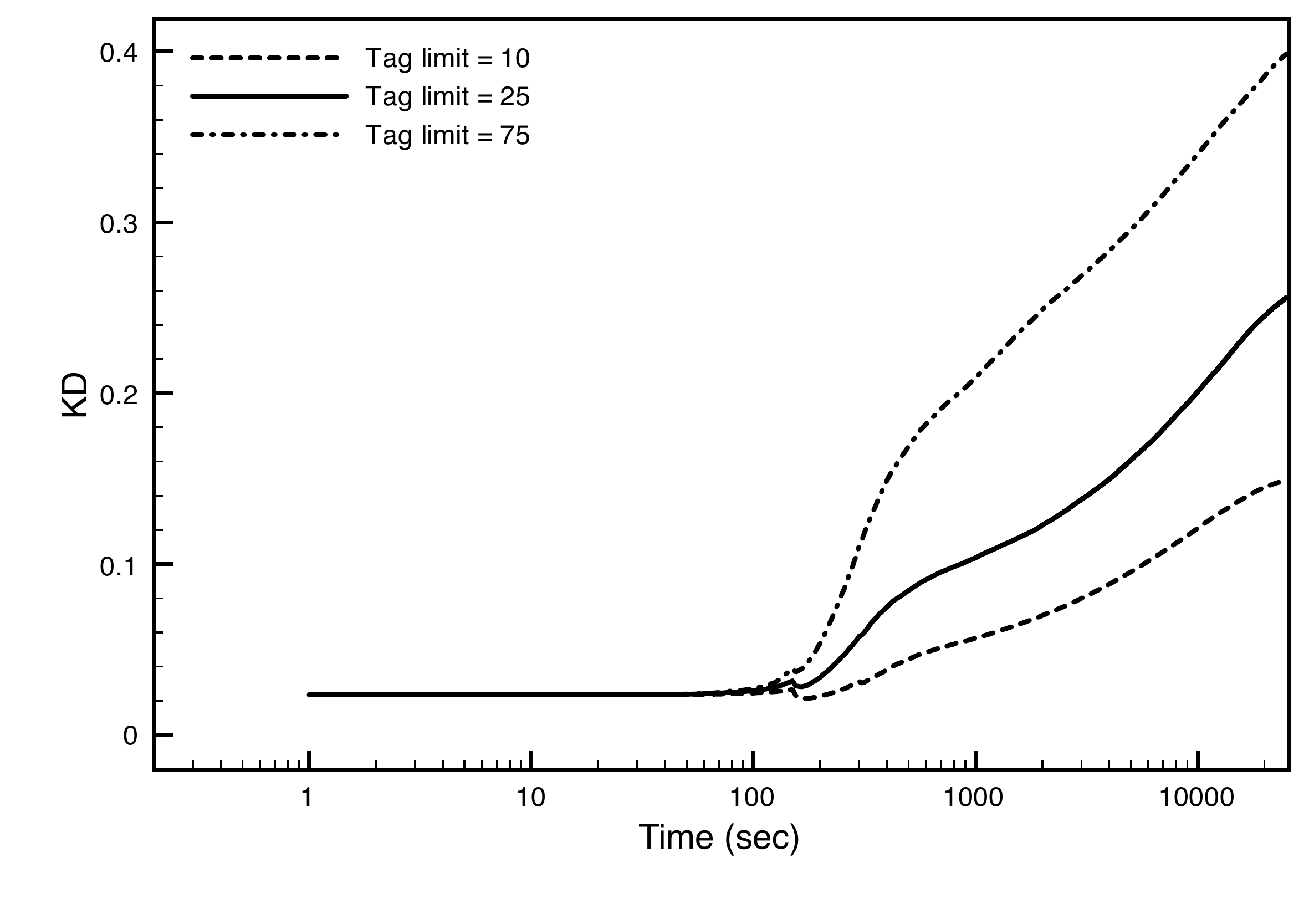}
  \caption{Impact on KD of the {\em tag\_limit} value} \label{fig:sensTags}
\end{figure}

Fig.~\ref{fig:sensTags} shows how the KD metric changes in front of different
values of {\em tag\_limit}, which sets the maximum number of SN tags 
exchanged during a contact. The inclusion of more vertices in a contributed
network leads to a significant increase of KD. Therefore, the setting of this parameter should be
  carefully evaluated, as increasing it also means increasing the amount of
  traffic exchanged during contacts. However, again we can notice that the
  increase in KD is not proportional to the increase \emph{tag\_limit}, which
suggests a marginal utility law also in this case.

\begin{figure}[H]
  \centering%
  \subfloat[{\em F-measure}]
  {\includegraphics[width=0.48\textwidth]{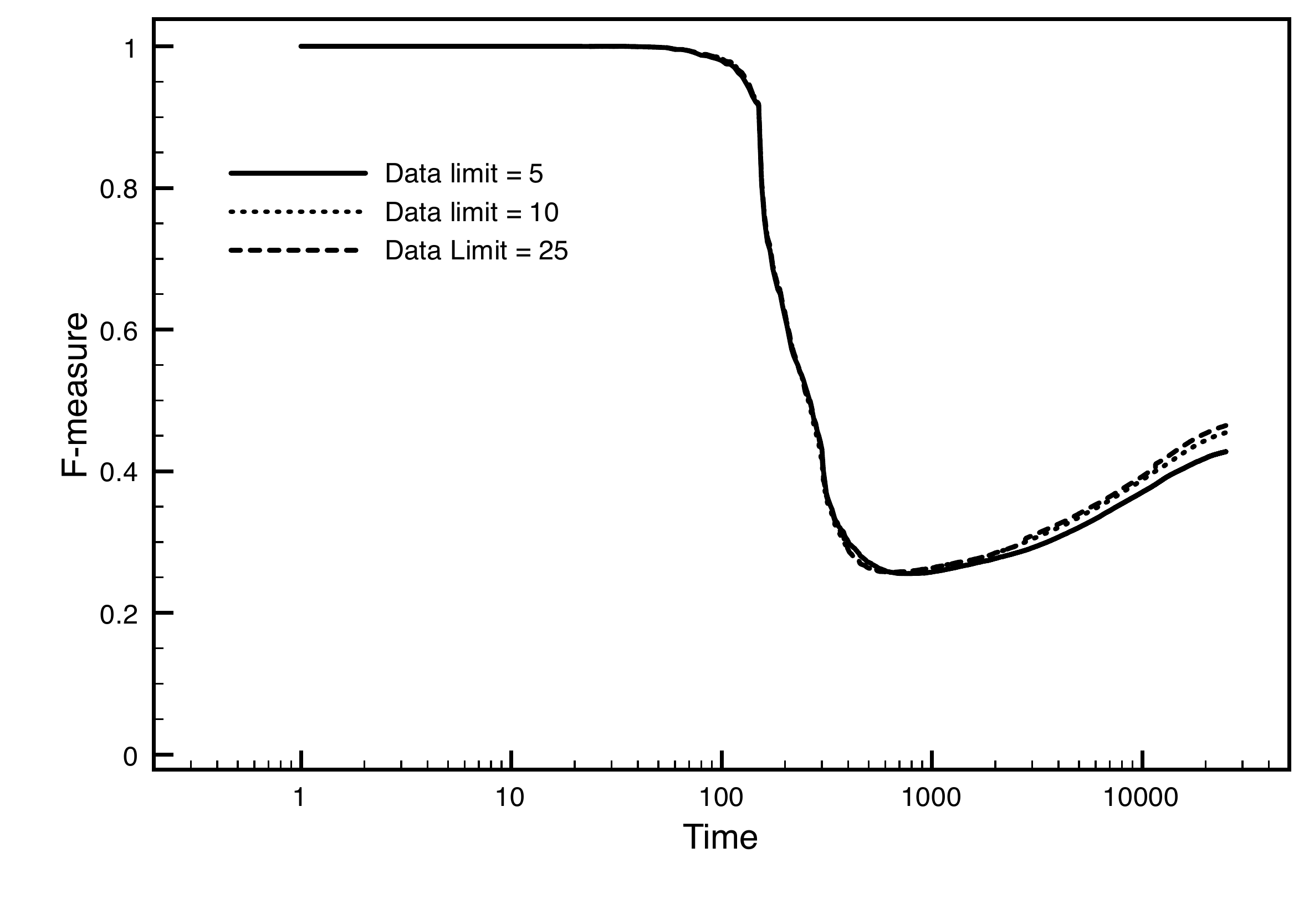}
    \label{fig:fmSenseDL}}
  \subfloat[Coverage]
  {\includegraphics[width=0.48\textwidth]{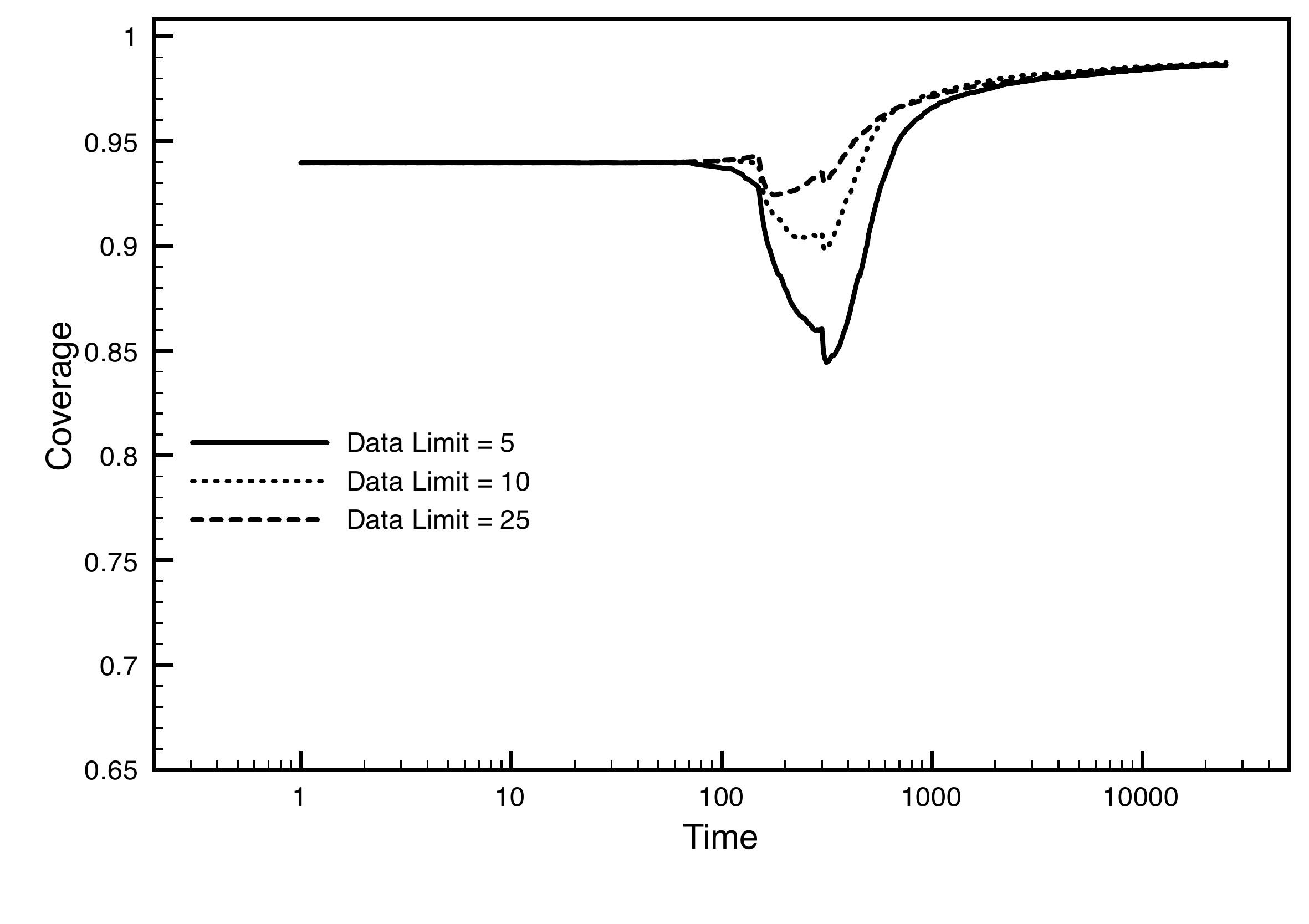}
  \label{fig:covSenseDL} }
  \caption{Impact of the {\em data\_limit} value on content dissemination
  \label{fig:sensCD}}
\end{figure}

Finally, Fig.~\ref{fig:sensCD} presents the impact on the content dissemination
process of various  {\em data\_limit} values, which limit the number of
content items exchanged during a contact. In particular,
Fig.~\ref{fig:fmSenseDL} presents the impact of this parameter on  F-measure
, while Fig.~\ref{fig:covSenseDL} shows its
impact on
Coverage. Note that an increase in the number of items

exchanged during contacts does not seem to particularly affect the value of the
F-measure. This can be explained by looking at how this metric is computed. One
of the main contributions in the computation of the F-measure value is given by
the intersection of all the semantic concepts that describe the received data
items and the semantic concepts stored  on the semantic network of a node.
Exchanging more items during contacts
could not lead to increase of the this quantity.
In fact, the items that maximise the intersection described above
are likely to be the first ones in the set considered for
exchange, because they are those most correlated with the concepts
exchanged during the same contact. Therefore, they
could be always present, even with the smallest amount of exchanged items (i.e.
{\em data\_limit} = 5 in our case).

On the other hand, when nodes can exchange (and, thus, also receive) more data
items at each encounter, they can more rapidly collect all the items that are
described by the tags they carry in their semantic networks. Therefore, the
Coverage measure achieves higher values, and converges more rapidly, when {\em
data\_limit} is set to higher values. Moreover, also in this case we can notice
a marginal utility behaviour. In fact, even increasing the resources involved in
the dissemination process, we obtain only a
marginal increase of performance. This means that we can save the resources of
mobile devices without degrading significantly the overall performance
of the system.  

	\section{Conclusion} \label{sec:conclusion}
In this paper we presented a set of algorithms enabling nodes belonging to an
opportunistic network to disseminate autonomously both content items and
semantic information associated to them. With respect to the work present in the
literature we are able to represent interests of users in a more flexible way
with respect, for example, to conventional topic-based schemes. Here, we
consider that users' interests can change over time, as a result of a knowledge
acquisition process that is also the effect of social interactions between them.
In a scenario of cyber-physical convergence, where users' devices are the main
tools through which users acquire knowledge and content from both the physical
and the cyber environment, this is important, because devices should act as much
as possible as proxies of the users in the cyber world, and therefore behave as
closely as possible as their users would behave in the same conditions. 
%their users contents in line with their current interests. 
To this end, we based our solution on well established models coming from the
cognitive science field. Namely, we exploited the associative network memory
model to build a representation of the semantic knowledge on which we run two
fast and frugal decision making strategies, i.e. Fluency and Tallying
Heuristics, in order to identify which semantic data and data items to exchange
between nodes upon contacts. Our simulation results demonstrate how the
proposed  cognitive solution can be a valuable approach to disseminate contents
in an opportunistic network.  We compared the performance of our approach with
an alternative that does not exploit cognitive models.  Interestingly, benefits are
manifold. First, the proposed cognitive solution is able to build a more stable
representation of the semantic information present in the environment. This
means that mobile devices spontaneously select the ``hot'' information
available
in the system during time and organise it in a robust structure permitting its
long lasting permanence in their memory. Second, nodes' internal semantic
knowledge representation  approximates very well, though at a reduced scale, the
structural  properties of the whole knowledge present in the
environment. Indeed,
the SN locally built by each node with our cognitive
mechanism have a very similar
degree distribution with respect to the graph representing the
entire knowledge present in the environment. 
%It is worth noting that any node is aware of the entire semantic information
%present in the system.  
Finally, our results show that the  cognitive approach
achieves a more accurate
 retrieval of contents. In fact, our system
does not loose the acquired information
about the ``hot'' topics present in the environment and  is
thus able to disseminate those contents that  become
interesting over time, always keeping high levels of coverage.
Therefore, with our approach personal users devices are able to properly
adapt to the  dynamic change of interests always providing to their users
the contents they need.

% MATTEO TODO: Marco suggerisce di tagliare un po' le autocitazioni. 

%in our solution nodes  are able to exploit the semantic information associated
%with contents own by a user to build a high level representation of their
%interest and use it to spread contents that satisfy those interest. Moreover,  

	%\vspace{-0.3em} ACKNOWLEDGMENT \begin{acks} This work is funded by the
	%EC under the FET AWARENESS RECOGNITION (FP7-257756), FIRE EINS
	%(FP7-288021).  \end{acks} \vspace{-1.0em}
	
	%
	%
	% trigger a \newpage just before the given reference number - used to
	% balance the columns on the last page adjust value as needed - may need
	% to be readjusted if the document is modified later
	%\IEEEtriggeratref{8} The "triggered" command can be changed if desired:
	%\IEEEtriggercmd{\enlargethispage{-5in}}
	%
	%
	%		REFERENCES
\section*{Acknowledgments}
This work was partially funded by the European Commission under the MOTO (FP7 317959) and EIT ICT Labs MOSES (Business Plan 2015) projects.	

\section*{References}
\bibliographystyle{elsarticle-num} 

\bibliography{bibliography}
	%
	% that's all folks
\end{document}